\documentclass[aps,pre,10pt,secnumarabic,twocolumn,amssymb,notitlepage,superscriptaddress]{revtex4-2}
\usepackage[utf8]{inputenc}
\usepackage{epsf,graphicx,amssymb,amsmath,color}
\usepackage{xcolor}
\usepackage[squaren, Gray, cdot]{SIunits}
\usepackage{hyperref}

\hypersetup{colorlinks=true,linkcolor=blue,citecolor=blue}

\DeclareMathOperator\arctanh{arctanh}

\begin{document}

\title{ Skin formation in evaporating colloidal droplets }

\author{Raphael Saiseau}\email{r.a.saiseau@utwente.nl}
\affiliation{Physics of Fluids Department, University of Twente, 7522 AE, Enschede, The Netherlands}

\author{Lorenzo Botto}
\affiliation{Process and Energy Department, ME Faculty of Mechanical Engineering, TU Delft, 2628 CD Delft,
The Netherlands}

\author{Christian Diddens}
\affiliation{Physics of Fluids Department, University of Twente, 7522 AE, Enschede, The Netherlands}

\author{Alvaro Marin}\email{a.marin@utwente.nl}
\affiliation{Physics of Fluids Department, University of Twente, 7522 AE, Enschede, The Netherlands}

\begin{abstract} 
When a droplet containing a concentrated suspension evaporates in a dry environment, a layer often forms at the interface accumulating non-volatile material. Such a ``skin layer'' experiences strong stresses and eventually turns mechanically unstable at the last stage of evaporation. 
Predicting the formation of such skin layer or particle shell and its properties is a crucial problem for applications and constitutes a multi-scale problem, from the micro/nanoscopic scale of the particles to the millimetric size of the droplets. Interestingly, its physical description lies at the interface between deterministic macroscopic evaporation models and microscopic stochastic particles interactions and diffusion.
In this work we present a general theoretical approach to obtain the time-dependent particle concentration profile in an implicit manner, for the general case of diffusion-limited evaporation of spherical droplets, and more generally to all 1D non linear diffusion-limited cases with particles pressure and mobility terms of rational form. This approach is compared successfully to numerical solutions obtained using a finite element solver in the limit of high P\'eclet numbers, and to 2D Brownian dynamics simulations.
Our results show that the concentration profiles and shell formation onset depend nontrivially on the initial packing fraction. 
By analyzing these profiles, we determine the position where the glassy layer forms, whose formation is expected to play a critical role in shell buckling. This model provides a robust framework for predicting the size and maximum aspect ratio of the resulting clusters.


\end{abstract}

\maketitle

\section{Introduction}\label{sec:intro}

When a volatile droplet containing a small concentration of colloidal particles is deposited on a regular flat substrate, the pinning of the contact line leads to an evaporation-driven flow. Such a flow brings the particles to the droplet‘s edge, ultimately resulting in a characteristic ring-shaped stain, the so-called ``coffee stain'' effect \cite{deegan1997capillary,gelderblom2022evaporation}.

However, when the droplet is deposited on a super-hydrophobic substrate with negligible contact angle hysteresis, the coffee-stain effect is now suppressed and the evaporation-driven flows arising within the droplet cannot accumulate particles in any specific location within the droplet. For strong super-hydrophobic states and high initial particle concentrations, the remains of the droplet are not shaped as a quasi-two-dimensional stain but rather as a three-dimensional particle cluster with almost spherical symmetry, often referred as ``supraparticle'' in the literature \cite{marin2012building,wooh2015synthesis,seyfert2021evaporation,Wang2024Supraparticle}. Such \emph{Supraparticles} are of practical interest since they can be used in cosmetics and food industry as units for powders or as micro-capsules \cite{Wang2024Supraparticle, wintzheimer2021supraparticles}.


However, when supraparticles are being formed following such evaporation process, a layer often forms at the interface accumulating non-volatile material.
Particles do not accumulate at the interface due to evaporation-driven flows (no streamline starts or ends at the liquid-air interface), but rather due to the receding liquid-air interface, which becomes very effective in sweeping colloidal particles along, generating a particle concentration gradient within the droplet \cite{gelderblom2022evaporation}. 
Such a ``skin layer'' or ``particle shell'' experiences strong stresses and eventually turns mechanically unstable at the last stage of evaporation. Unstable particle shells often experience buckling, which often ruins the supraparticle structure. Therefore, being able to predict the time-dependent particle concentration profile is of crucial importance since it allows to predict the onset of skin layer formation. 

This problem has been confronted by others in the past. From the experimental side, Seyfert \emph{et al.} \cite{seyfert2021evaporation} showed the important role of the initial particle concentration in mechanical stability of the resulting supraparticle in super-hydrophobic substrates, and revealed a minor role of the ambient humidity in the typical range achieved in open or non-pressurized systems.
Similar results have been found in hydrophobic substrates \cite{basu2016towards}, levitating Leidenfrost droplets \cite{tsapis2005onset}, drying droplets in electrodynamic levitation \cite{archer2020drying}, spray drying \cite{lintingre2016control} or disk-shaped drying \cite{boulogne2013buckling,bouchaudy2019drying}. 

From the numerical and theoretical side, Sobac \emph{et al.} \cite{sobac2019mathematical} developed a mathematical model that could be solved numerically to obtain the time-dependent  concentration profile in an evaporating spherical droplet containing colloidal particles. One of the strongest points of this model is the possibility of computing mechanical stresses at the particle shell, based on Darcy’s law of pressure gradients in porous media, crucial for predicting the stability of the supraparticle. A similar approach was followed by Daubersies \& Salmon \cite{daubersies2011evaporation}, but using a 2D geometry, which lead to further disk-shaped particle-laden droplet evaporation experiments \cite{loussert2016drying,sobac2020collective}.

Previous theoretical and numerical approaches to the problem aimed to understand the role of the P\'eclet number and compute the time-dependent collective diffusion within the compacted region \cite{,daubersies2011evaporation,sobac2020collective}. In this work, we aim to elucidate the role of the initial particle packing fraction and the inter-particle interactions in the particle shell formation.

To achieve this, we develop an analytical model in which the non-linear diffusion problem is solved in a moving reference in the limit of high P\'eclet numbers and use it to theoretically investigate the role of the initial packing fraction. We show that this approach enables to implicitly solve this problem for all rational equation of state and mobility terms.\\ 

The paper is organized as follows, in section \ref{sec:intro_model} we introduce the general framework that will be used along the paper. In section \ref{sec:quasi-static}, we compare the quasi-static limit solution to a numerical solution of the diffusion equation using a finite-element solver. 
The model is then tested comparing the results with 2D Brownian dynamics particle simulations in section \ref{sec:simulations}. To show the flexibility of the model, the model is applied to the case of particles interacting as hard spheres in section \ref{sec:hard spheres}. The paper is concluded with a general conclusion and final discussions in section \ref{sec:conclusion}. 

\section{Theoretical model for a shrinking Brownian confined suspension}\label{sec:intro_model}

\begin{figure}
     \includegraphics[width =.85\columnwidth]{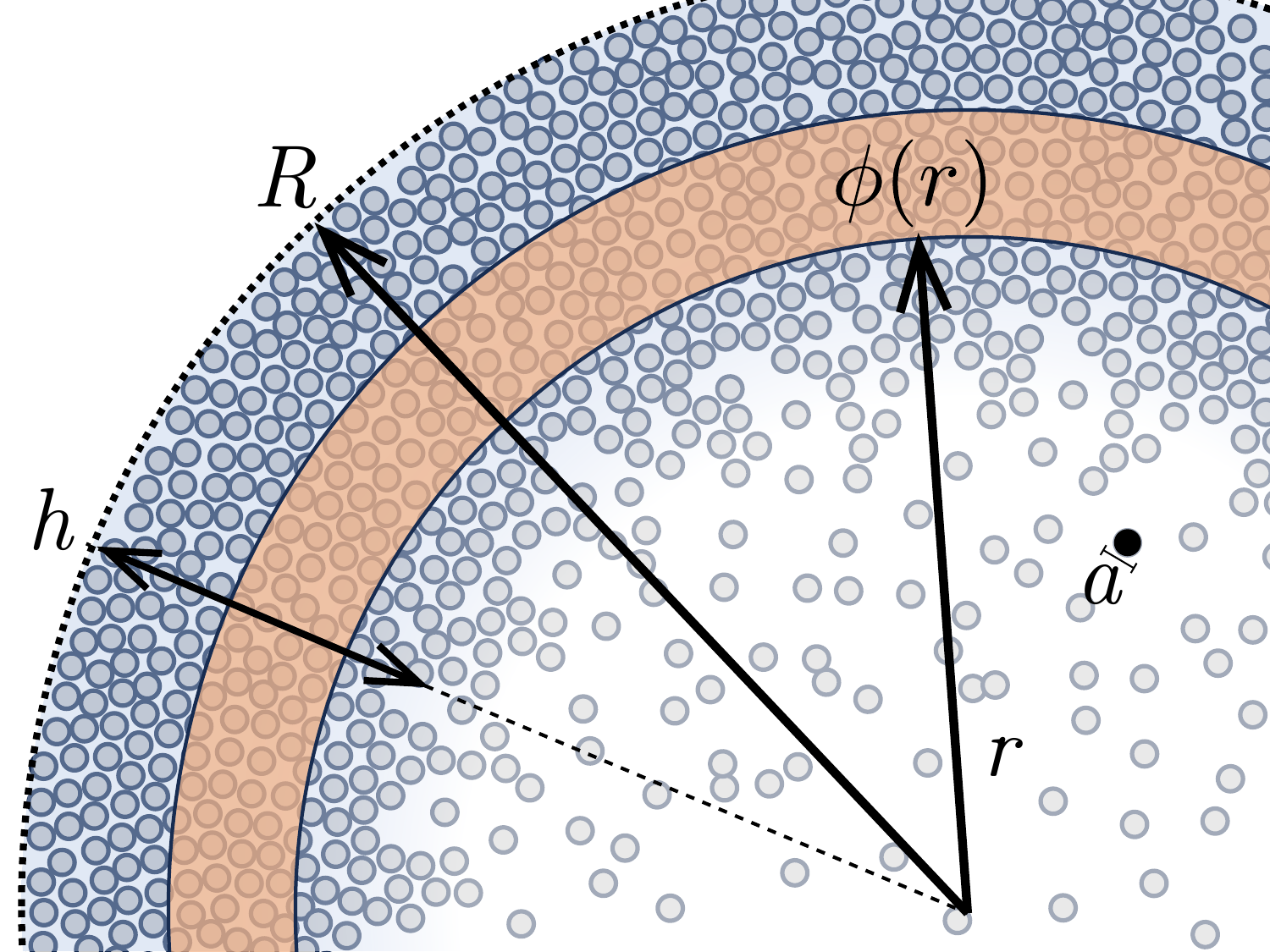}
\caption{Scheme describing the main parameters in the theoretical model in section \ref{sec:intro_model}. The scheme shows a shrinking droplet of radius $R$ containing Brownian particles of diameter $a$, at an advanced stage of the evaporation process, where a particle shell of thickness $h$ has been formed.}
\label{fig:Scheme}
\end{figure}

We consider a Brownian suspension of monodisperse particles particles of diameter $a$ confined within a spherical domain of radius $R$ (see scheme in Figure \ref{fig:Scheme}). The initial particle distribution is homogeneous with initial volume fraction $\phi = \phi_0$.  The continuity equation for the particle phase is \cite{drew1982mathematical,zhang1997momentum}:

\begin{equation}
\frac{\partial \phi}{\partial t} + \nabla \cdot \left(\phi \mathbf{v}_p \right ) = 0,
\end{equation}

where $\mathbf{v}_p$ is the average particle velocity. The suspension is confined within an spherical droplet that evaporates in a diffusion-limited regime with homogeneous evaporation flux at its interface and in thermal equilibrium, therefore no liquid flow appears in the droplet \cite{gelderblom2022evaporation}. However, the kinematic boundary condition at the receding liquid-air interface requires the particle velocity to be equal to the interfacial velocity for those particles at the interface, hence $\mathbf{v}_p = \dot{R} \mathbf{e}_r$ at $r=R$, with $\mathbf{e}_r$ the radial unit vector. 

Calculating $\mathbf{v}_p$ requires the solution of the coupled set of momentum and continuity equations for the particle and fluid phases \cite{drew1982mathematical}. However, a simple evolution equation for the particle volume fraction can be found by using the steady low-Reynolds-number momentum equation for the particle equation and a simple closure for the particle drag.  Neglecting shear components in the particle pressure $\Pi$, the particle momentum equation reads  

\begin{equation}
0 = - \nabla \Pi + \mathbf{f}^{hd},
\label{Eq:fluidmomentum}
\end{equation}

where $\mathbf{f}^{hd}$ is the drag force on the particles per unit volume of mixture. The particles pressure is written here like $\Pi = k_B T n Z(\phi)$ where $k_B$ is the Boltzmann constant, $T$ the system temperature and $n= \phi/ (\pi/6  a^3)$ is the particle number density. $Z(\phi)$ is a compressibility term taking into account particles interactions. For infinitely diluted system, $Z(\phi)\rightarrow 1$ such that the particles system reduces to a ideal gas of particles.
The drag force exerted by the fluid on a single spherical particle of diameter $a$ immersed in a fluid of viscosity $\mu$ is $3\pi \mu a (\mathbf{v}_f - \mathbf{v}_p)$, where $\mathbf{v}_f$ is the volume-averaged fluid velocity. Thus, for $\phi \rightarrow 0$ the volumetric drag density is

\begin{equation}
\mathbf{f}^{hd} = \xi_r n (\mathbf{v}_f - \mathbf{v}_p)  
\label{Eq:drag}
\end{equation}

where the resistance coefficient is $\xi_f=3\pi \mu a / f(\phi)$ with $f(\phi) \rightarrow 1$ for an infinitely diluted medium. Using equations (\ref{Eq:drag}) and (\ref{Eq:fluidmomentum}),  the particle velocity can be expressed in terms of the fluid velocity and gradient of particle pressure. If we assume that the particle pressure is only a function of $\phi$, then  the continuity equation for the particle phase is a closed equation in $\phi$ and is thus solvable if $v_f$ is known. This approach is at the basis of the suspension balance model for particle suspensions, as discussed e.g. in Ref. \cite{nott2011suspension}.  

The approach is less straightforward for a dense suspension, because in that case the drag force can be a function of particle concentration and particle microstructure \cite{squires2010fluid}. Here this effect is considered through the term $f(\phi)$ in $\xi_r$. The chosen expression of $\xi_r$ will be discussed later.

Since our system is a spherical droplet shrinking its diameter due to diffusion-limited evaporation, there not flow is induced within the droplet \cite{gelderblom2022evaporation} and $\mathbf{v}_f = 0$. 
 Introducing the mobility coefficient $\lambda = (\pi/6  a^3)/ \xi_r$ we get

\begin{equation}
\frac{\partial \phi}{\partial t} = \nabla \cdot \left[\lambda \nabla \Pi \right]
\label{Eq:Gen_eq}
\end{equation}

Note that $\Pi$ is often regarded as an osmotic pressure, however the concept of particle pressure is in general more appropriate for a system out of equilibrium as a particle-laden evaporating drop, \cite{deboeuf2009particle}. 

Evaluating the particle momentum equation at $r~=~R$ gives  $\mathbf{v}_p(R) \phi(R) = - \lambda \partial_r \Pi(R)$. Using the kinematic boundary condition for the particle velocity $\mathbf{v}_p = \dot{R} \mathbf{e}_r$, this becomes  
\begin{equation}
\lambda \partial_{r} \Pi \vert_{R} = - \phi(R) \dot{R}
\label{Eq:kin_BC}
\end{equation}

Finally, to comply with the system symmetry, a no flux
boundary condition needs to be invoked at the domain center:
\begin{equation}
\partial_{r} \Pi \vert_{r=0} = 0
\label{Eq:symm_BC}
\end{equation}

In the following, a numerical solution of such system of equations will be computed using a finite element framework \cite{diddens2017detailed,diddens2017evaporating}.

\subsection{Quasi-static solution}\label{sec:quasi-static}

To illustrate the resolution process and highlight the main physical contributions to the solutions, we solve here the problem in the absence of hydrodynamic interactions in a quasi-static limit, i.e. dropping the left-hand side of Eq. \ref{Eq:Gen_eq}. The particles self-diffusion coefficient is given by $D_0 = \lambda k_B T$ with the mobility coefficient $\lambda$ being here independent of the $\phi$. The compressibility term $Z(\phi)$ is given here in a rational form \cite{routh2004distribution}:
\begin{equation}
Z(\phi) = \left( \frac{\phi_m}{\phi_m - \phi} \right)^{\alpha}
\label{Eq:Comp_term}
\end{equation}
with $\phi_m$ is the maximum packing fraction which is chosen either from the random close packing value $\phi_{\mathrm{rcp}}$ or the maximum packing packing $\phi_{\mathrm{mp}}$. We show here the case $\alpha = 1$ corresponding to diverging contribution for a Van der Waals gas. The solution for any $\alpha$ can be found in the Appendix \ref{Appendix:generalsolution}. \\


In order to simplify the formulation in a non-stationary shrinking system, here we assume that there exists a moving reference frame, where the problem is quasi-static, i.e. where $\partial \phi/\partial t = 0$. To take into account the radial symmetry of the problem, the moving reference frame coordinates is given by $r'=r-\int^t_0 u(r,t') dt'$ where $u$ is the relative fluid velocity observed from the moving reference frame. In order to fulfill mass conservation on a elementary volume for a homogeneous distribution of particles of packing fraction $\phi$, 

\begin{equation}
\nabla (u \phi) = 0  \rightarrow \nabla u= 0,
\end{equation}

which is practically identical to an incompressibility condition and satisfies $\nabla_{r'} = \nabla$. This yields $u = {U(t)}\left(\frac{R}{r} \right)^{n-1}$, where $U(t)$ is a characteristic velocity which does not depend on the radial coordinate but proportional to $\dot{R}$, and $n$ is the system’s dimension. Note that $u$ is not a real flow field and therefore it does not satisfy Navier-Stokes equation.

Considering the boundary conditions \eqref{Eq:kin_BC} at the droplet center and the air-water interface, equation \eqref{Eq:Gen_eq} then reduces to:
\begin{equation}
     - \dot{R}\frac{\phi(R)}{\phi(R) - \phi(0)} \left(\frac{R}{r} \right)^{n-1} (\phi-\phi(0)) = D_0 \nabla \left[\phi Z(\phi) \right]
\label{Eq:const_eq}
\end{equation}

In order to solve for $\phi(r)$, we need to define two boundary conditions. In this quasi-static solution, we will assume $\phi(0) = \phi_0$, where $\phi_0$ is the initial particle concentration in the droplet from the laboratory frame of reference. Regarding the particle packing fraction at the interface $\phi(R)$, we will assume it is the maximum packing packing fraction allowed in the system $\phi_m$, which value depends on particle interactions and will be discussed further below. 




For the dilute case scenario $Z(\phi) \rightarrow 1$, the concentration profile builds up at the water-air interface with an exponential profile as $ e^{ (R-r)/ \xi}$ with $\xi = \frac{\phi(R) - \phi_0}{\phi(R)} D_0 / \vert\dot{R}\vert$, the characteristic length scale for the concentration profile. 
The condition for the concentration profile to build up sufficiently for the shell to form is then given by $\xi \ll R$ which corresponds $\mathrm{Pe} \gg 1$ with $\mathrm{Pe} = R \vert \dot{R} \vert / D_0$ a P\'eclet number. 
In diffusion-limited droplet evaporation, a droplet typically follows the well-known d2-law \cite{Langmuir:1918}, such that $R^2 \propto t$, and consequently $\dot{R} \propto 1/R$. This leads to a time-independent $\mathrm{Pe}$, a very  convenient control parameter to determine the onset of shell formation.
We stress here that the approach of choosing a reference frame where the problem can be considered as quasi-static in a way to include the boundary conditions does not have to be restricted to this problem \emph{a priori}. 

\begin{figure}[ht!]
     \includegraphics[width=.85\columnwidth]{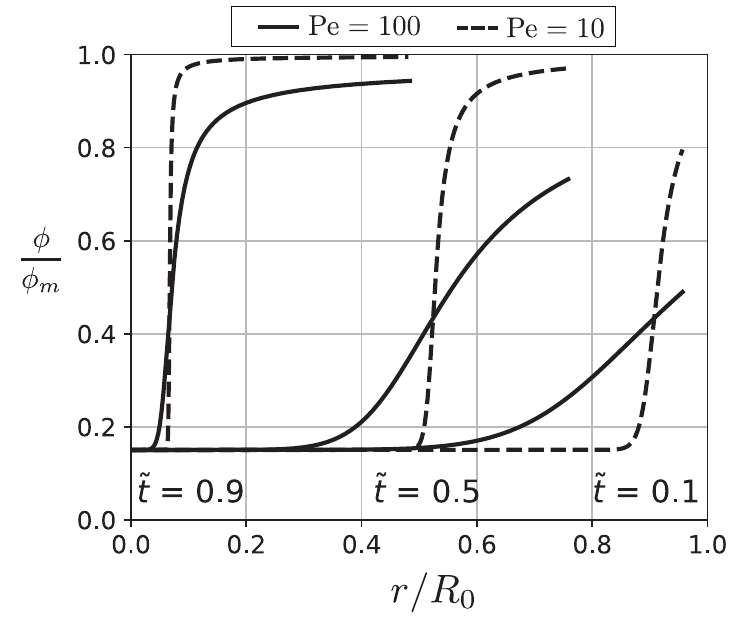}
    \caption{Particles volume fraction profiles at varying times with $\tilde{t}= t / \tau_{\mathrm{min}}$ with $\tau_{\mathrm{min}} = (R^2_0-R^2_{min}) / 2 \mathrm{Pe} D_0$, obtained from Eq. \eqref{Eq:Sol_alpha1} for circular geometry.}
\label{fig:Sol_2D}
\end{figure}

\begin{figure*}
     \includegraphics[width=.8\columnwidth]{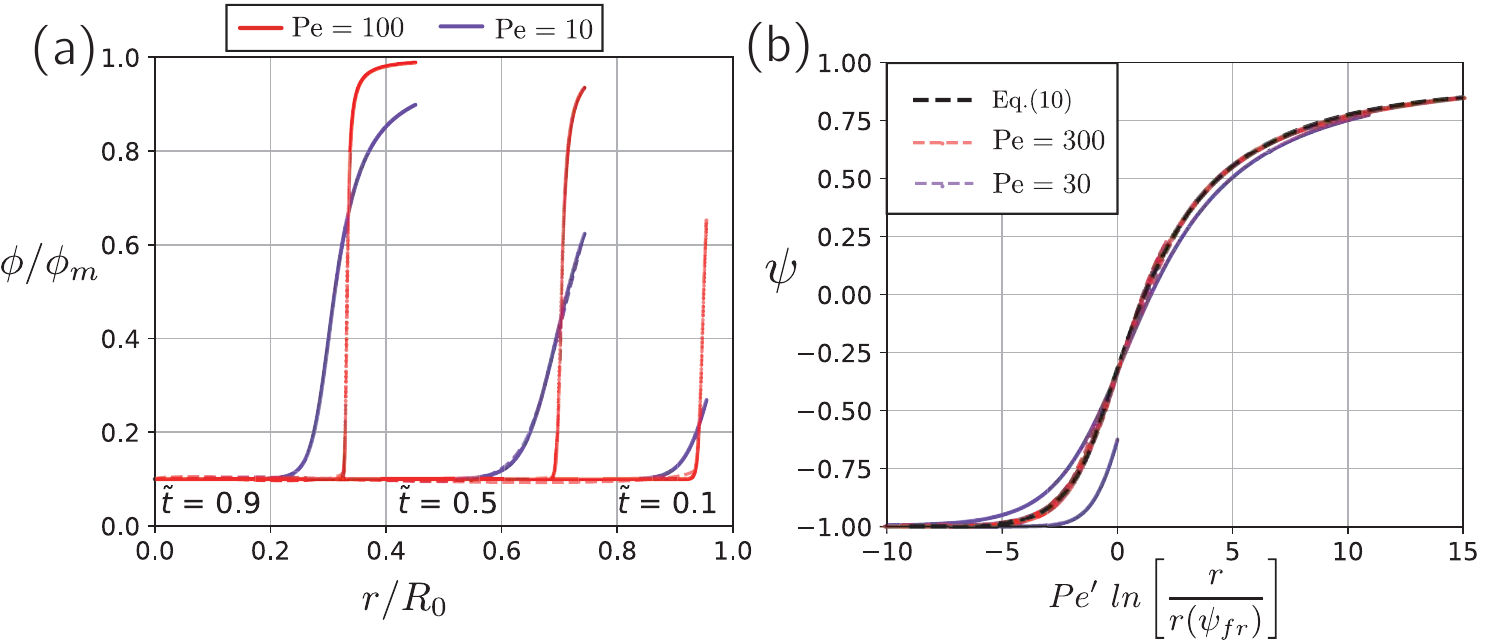}
     
    \caption{(a) Numerical packing fraction profiles for $\alpha = 1$ and circular geometry at varying times $\tilde{t}= t / \tau_{\mathrm{min}}$ with $\tau_{\mathrm{min}} = (R^2_0-R^2_{min}) / 2 \mathrm{Pe} D_0$. (b) Rescaled profiles compared to equation \eqref{Eq:Sol_alpha1}.}
\label{fig:Sol_2D_comp}
\end{figure*}

For $\mathrm{Pe} \gg 1$, as the shell to form, the concentration profile at the water-air interface saturates giving $\phi(R) \approx \phi_m$ the maximum packing fraction. Here we rescale the volume fraction into a new variable $\psi$, defined as
\begin{equation}
 \psi = 2\frac{\phi-\phi_0}{\phi_m-\phi_0} - 1,
 \label{eq:psi}
\end{equation}

which allows us to solve equation \eqref{Eq:const_eq} by separating variables and using simple element decomposition. For $\alpha = 1$, it yields:

\begin{equation}
\left[\arctanh{\left( \psi \right)} + \left( \frac{1}{1-\psi} \right)\right]^{\psi}_{\psi_{\mathrm{fr}}} = \frac{\mathrm{Pe'}}{2} \left[g\left(\frac{r}{R}\right) \right]^{r}_{R-h}
\label{Eq:Sol_alpha1}
\end{equation}

where $\mathrm{Pe'} = \frac{\phi_m - \phi_0}{\phi_m} \mathrm{Pe}$ and $g$ is a function which depends on the system’s dimension:


\begin{align*}
    g\left(\frac{r}{R}\right) &= \frac{r}{R}, &\quad\text{for 1D},\\
    g\left(\frac{r}{R}\right) &= \ln\left( \frac{r}{R}\right), &\quad\text{for 2D},\\
    g\left(\frac{r}{R}\right) &= \frac{R}{r}, &\quad\text{for 3D},\\
\end{align*}

corresponding to a symmetric linear system (1D), a disk (2D) or a sphere (3D). 

We define the shell front position $r(\psi = \psi_{\mathrm{fr}})$ as the position of the maximum gradient in $\psi(r)$, where it takes the value $\psi_{\mathrm{fr}}$, obtained from equation \eqref{Eq:const_eq}. As equation \eqref{Eq:Sol_alpha1} can not be integrated directly, $h = R - r(\psi = \psi_{\mathrm{fr}})$ the shell thickness is obtained supposing an ideal bulk-shell structure, with two distinct phases, a bulk of packing fraction $\phi_0$ and a shell of packing fraction $\phi_m$ \cite{daubersies2011evaporation,boulogne2013buckling,sobac2019mathematical,milani2023double}:
\begin{equation}
    \frac{h}{R} = 1- \left\lbrace \frac{\phi_m}{\phi_m-\phi_0} \left[1 - \left(\frac{R_{min}}{R}\right)^{n} \right] \right\rbrace^{1/n}
\label{Eq:h}
\end{equation}

Note that we define an end radius $R_{min}$, where $h/R_{min}=1$ and the bulk-shell assumption breaks down. This takes the value $R_{min} = \left( \phi_0 / \phi_m \right)^{1/n} R_0$. Accordingly, equation \eqref{Eq:h} for $h$ converges to 0 at $R_0$.

Figure \ref{fig:Sol_2D} shows the analytically obtained solution \eqref{Eq:Sol_alpha1} for the 2D case for different P\'eclet numbers and Fig. \ref{fig:Sol_2D_comp} shows its comparison to the numerical solutions of Eq. \ref{Eq:Gen_eq}, with boundary conditions \ref{Eq:kin_BC} and \ref{Eq:symm_BC}. The solution separates the space in two regions with a transient region occupying a finite region of space. Using a Ginzburg-Landau model, the liquid-gas phase separating interface is given by a hyperbolic tangent profile in equation \eqref{Eq:Sol_alpha1} and imposes the intrinsic length scale $\xi$, here fixed by the P\'eclet number, to the solution. This front profile is expected for any left-hand term in equation \eqref{Eq:const_eq} which increases proportional to $\phi$ and is acting as an effective attractive interaction term. Here it comes from the problem being quasi-static in the moving reference frame and physically results from the transport of the particles by the droplet shrinking. 

The second polynomial term makes the profile asymmetric, going only asymptotically to $\psi = +1$ underlying the incompressible nature of the condensed phase of the profile.

Note that the solution is parametrized using time- and space-dependent coordinates, and therefore it propagates both in space and time. This characteristic originates from the growth of the condensed region here parametrized by equation \eqref{Eq:h}. Since $\xi \sim R /\mathrm{Pe}$, as the droplet evaporates, the front profile also becomes steeper with time (see Figure \ref{fig:Sol_2D}). This result would then be a generic feature of this problem and should not depend on the choice of the equation of state or mobility term while the regime stays quasi-static.

We considered here $\alpha = 1$ for the sake of simplicity but these conclusions also apply for higher $\alpha$ giving only supplementary polynomial terms in $\phi$ in equation \eqref{Eq:Sol_alpha1}. The general solution for any $\alpha$ is shown in the supplementary material. Taking into account hydrodynamic interaction in the form of a friction coefficient $f(\phi) = (1-\phi)^{6}$ however plays the opposite role as we will see later. The concentration profile becomes steeper close to the particle condensed region and smoother in the diluted region \cite{routh2004distribution}. 

\section{Brownian dynamics simulation}\label{sec:simulations}

The approach followed in the model described above assumes a continuum concentration profile, even though the suspension is composed of discrete particles. In this section we compare the results of the continuum model with discrete 2D Brownian dynamics simulations. 



\begin{figure}
\includegraphics[width =.85\columnwidth]{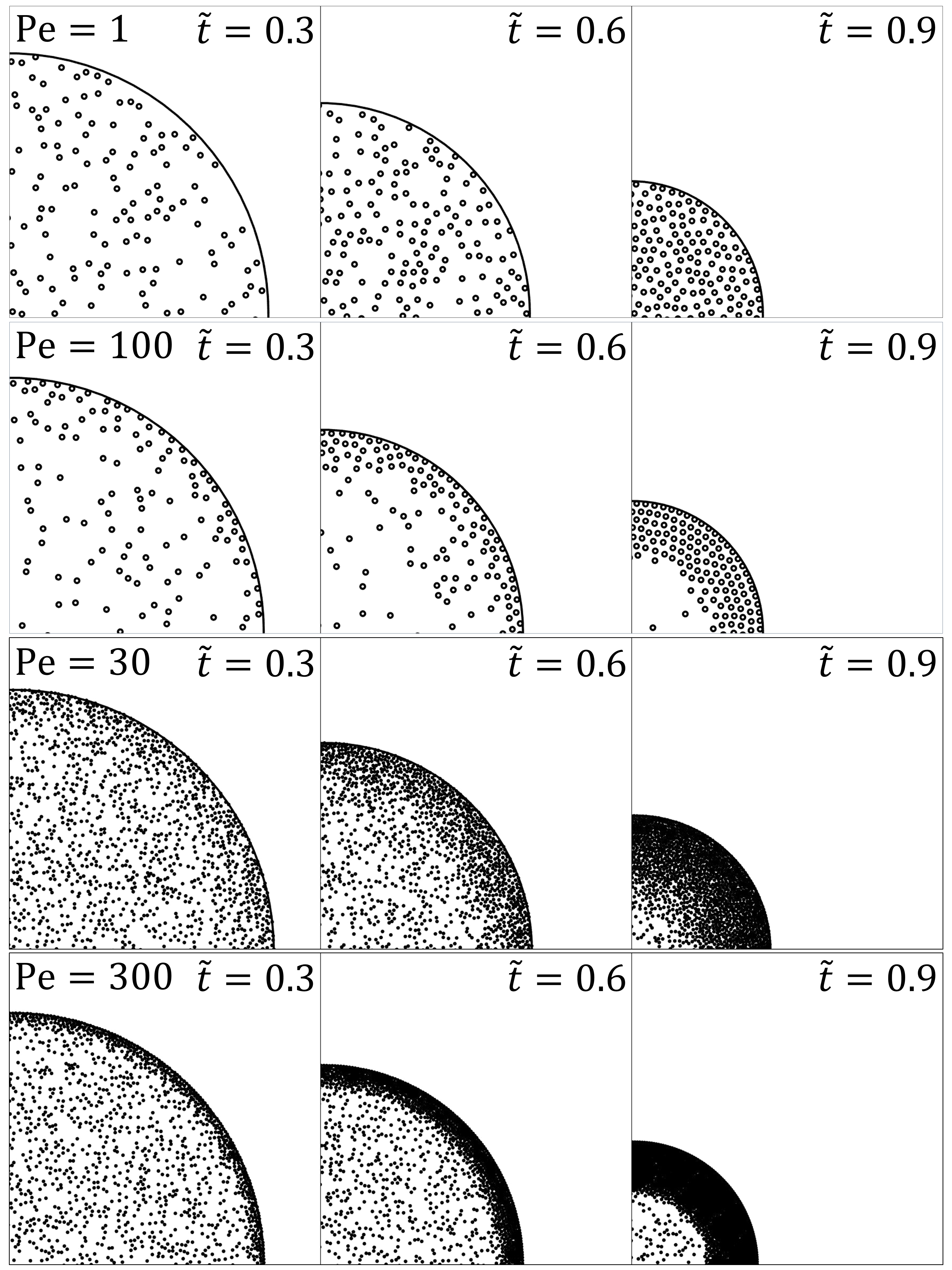}
\caption{Brownian particle dynamics simulations, snapshots at different times of the shrinkage process $\tilde{t}= t / \tau_{\mathrm{min}}$ with $\tau_{\mathrm{min}} = (R^2_0-R^2_{min}) / 2 \mathrm{Pe} D_0$, for 500 particles and $\mathrm{Pe} = 1$  and $100$, and for 5000 particles for $\mathrm{Pe} = 30$  and  $500$. Note that, given the lack of attraction among the particles, the simulation runs until all particles are compressed at their maximum packing fraction. This is clearly visible in the last snapshots of (a), (b) and (c).}
\label{fig:SimuImages}
\end{figure}
 
In the simulations we introduce a system of particles obeying an overdamped Langevin equation with a modified Weeks-Chandler-Andersen (WCA) interaction potential, which is a short-range repulsion identical to the repulsive short-range profile of the Lennard-Jones potential \cite{ladiges2021discrete}. The particles are initially distributed randomly in a circle domain that shrinks following a classical R2-law \cite{Langmuir:1918}, typical for diffusion-limited evaporation. Particles interact with the interface through a soft-boundary condition: when a particle attempts to walk out the interface, its position is corrected to the interface and its velocity component normal to the interface is canceled out. 
Note that this boundary condition is not ``adsorbent'', i.e. particles are not attached to the interface and they are always free to roam within the shrinking disk. In this sense, this condition is perfectly inelastic, but not adsorbent.
Nevertheless, a particle reaching the interface will be able to roam away from it only if it moves fast enough. This rivalry between the interface's and the particle's characteristic time scales is reflected in the P\'eclet number $\mathrm{Pe} = R \dot{R} / D_0$ with $D_0$ being here the particles associated self-diffusion constant driving the noise term of the Langevin equation.

\begin{figure}
     \includegraphics[width=.85\columnwidth]{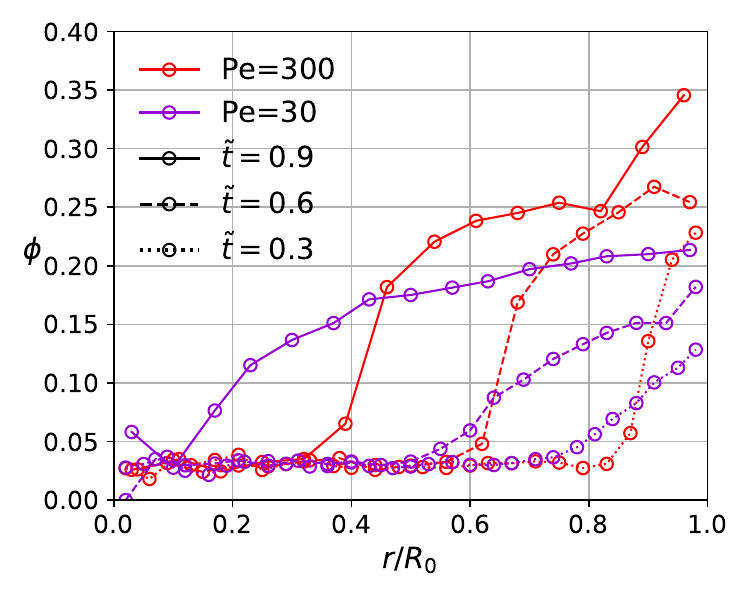}
    \caption{ Particles packing fraction profiles for Brownian dynamics simulation in shrinking domain at adimensionalised time $\tilde{t}=0.9,0.6,0.3$ with $\tilde{t}= t / \tau_{\mathrm{min}}$ with $\tau_{\mathrm{min}} = (R^2_0-R^2_{min}) / 2 \mathrm{Pe} D_0$.The initial packing fraction $\phi_{ini} = 0.03$ and the number of particles $N=5000$.}
\label{fig:Partsimu_profiles}
\end{figure}

Figures \ref{fig:SimuImages} show different time series respectively for 500 to 5000 particles inside a circle domain with P\'eclet numbers varying between 1 and 500. 
For $\mathrm{Pe} = 1$, the particles distribution shows an almost homogeneous profile with a concentration slowly growing over time until reaching a condensed phase where the particles are hexagonally ordered up to geometry related defects. However for moderate and high P\'eclet number ($\mathrm{Pe} = 30-100$), the particles distribution shows three phases: first, the accumulation of particles at the domain walls, then it develops towards two separated regions, and eventually the outer region takes over the full droplet. These two separated regions consist of a shrinking inner diluted region with a concentration close to its initial value $\phi_{ini}$, and a growing outer one, highly condensed and ordered (see Figure \ref{fig:SimuImages}). As the domain shrinks or as $\mathrm{Pe}$ increases, the transient zone connecting the two regions becomes steeper (see Figure \ref{fig:Partsimu_profiles}). For $\mathrm{Pe} = 500$ the transient zone is not observable anymore and an almost flat front separates both regions.

Qualitatively, these results compare well with those obtained in the quasi-static solution \eqref{Eq:Sol_alpha1} of the continuous model. To make a more quantitative comparison, in Figure \ref{fig:Partsimu_profiles} we show the particles packing fraction profiles $\phi = c \pi (a/2)^2$ with $c$ the local particle concentration, for which it is necessary to consider $a$ the particle diameter from the WCA interaction reach.
Note that the profiles tend to asymptotically converge to higher packing fractions than $\phi_m$ as $\mathrm{Pe}$ increases. This is a result of the continuous nature of the WCA interaction being unable to reproduce a hard core potential in a highly packed system under pressure from the shrinking environment. However, using this definition, the pressure of the particles is well approximated using a compressibility term $Z(\phi)$ with $\alpha = 2$ (see Appendix \ref{Appendix:generalsolution}).

\begin{figure}
     \includegraphics[width=.85\columnwidth]{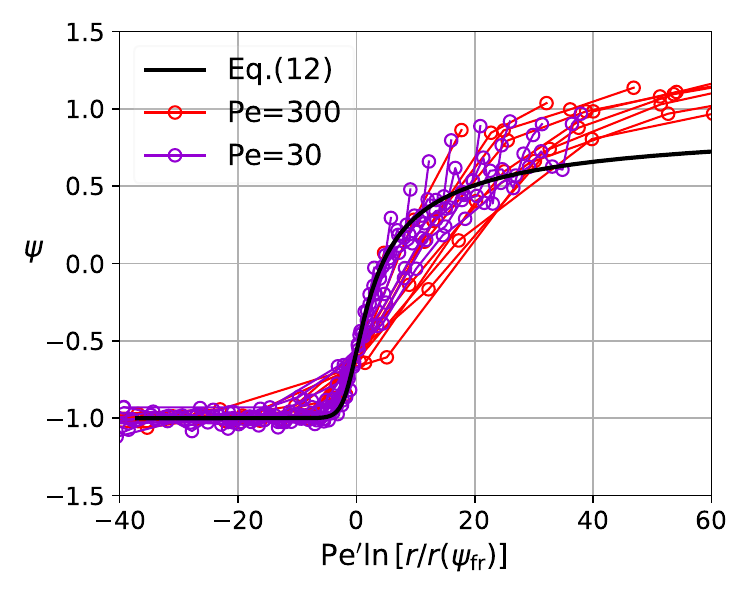}
    \caption{ Rescaled packing fraction profiles in late shell formation regime $h/R>2/\mathrm{Pe'}$ compared to Eq.\eqref{Eq:Sol_partsimu}.  The number of particles $N=5000$ and $\phi$ is defined elementary rings of width $2 \mathrm{d}r = 5 a$.}
\label{fig:Partsimu_profrescal}
\end{figure}

The quasi-static solution for $\alpha = 2$ in circular geometry is given by:
\begin{multline}
    \mathrm{Pe'} \ln{\left( \frac{r}{R-h} \right) } = \left\lbrace 4\left( \frac{1}{1-\psi}\right)^{2} - 4\left( \frac{1}{1-\psi_{\mathrm{fr}}}\right)^{2} +  \right. \\ \left.2\left( 1+\frac{\phi_0}{\phi_m} \right)  \left[ \arctanh(\psi)  + \frac{1}{1-\psi}\right] \right\rbrace^{\psi}_{\psi_{\mathrm{fr}}}
    \label{Eq:Sol_partsimu}
\end{multline}
with $\mathrm{Pe'} = ((\phi_m - \phi_0)/\phi_m)^{2} \mathrm{Pe}$ and $\psi_{\mathrm{fr}}= -0.57$. Figure \ref{fig:Partsimu_profrescal} compares this solution with the packing fraction profiles obtained from the Brownian dynamics simulations for $\mathrm{Pe} = 30$ and $300$.
We observe that the shape of the packing fraction profile is well reproduced by the solution in Eq. \ref{Eq:Sol_partsimu}. In particular, the predicted rescaling of the shell front is well retrieved as well as the asymmetric nature of the concentration profile.
Due to the second-order compressibility term, it also shows that the volume fraction approaches the maximum value at particularly long distances far from being reached by these simulations. 
The divergence of the simulations from Eq. \ref{Eq:Sol_partsimu} in the shell region (right side of the plot in Fig. \ref{fig:Partsimu_profrescal}) results from the soft nature of the system, as discussed above. 
Accordingly to what was observed for the comparison with the finite element numerical solution, this hints towards that the maximum packing fraction is hard to approach on corresponding physical systems of hard disks. This last point also leads Eq. \ref{Eq:h} to underestimate the shell front position (see Appendix \ref{Appendix:generalsolution}).
 
\section{Particular case: hard spheres suspension in liquid}\label{sec:hard spheres}


To demonstrate the versatility of the model proposed, we now consider a more traditional particle interaction model, a hard sphere model, closer to those proposed in references \cite{daubersies2011evaporation,sobac2019mathematical}. We use the Van der Waals mean field approximation for concentrated solution $Z(\phi) = 3/(1-\phi/\phi_m)$ generally used for hard spheres systems with $\phi >0.55$ \cite{buzzaccaro2007sticky,piazza2014settled,daanoun1994van} and add the Richardson-Zaki \cite{richardson1954sedimentation} hydrodynamic interaction term $f(\phi)=(1-\phi)^{6}$ to the mobility term $\lambda \rightarrow \lambda f(\phi)$  \cite{peppin2006solidification,russel1991colloidal}. The constitutive equation \eqref{Eq:const_eq} then reads for $\phi(R) \approx \phi_m$:
\begin{equation}
    \frac{\mathrm{Pe'}}{12(1-\phi_0)^6} \frac{1}{r'^{2}} =   \frac{\left[1-\frac{1}{\mathrm{Hy}}(\psi+1)\right]^{6}}{(1-\psi)(1-\psi^2)} \nabla' \psi
    \label{Eq:const_Hydro}
\end{equation}
with $\mathrm{Pe'} = \frac{\phi_m - \phi_0}{\phi_m}\mathrm{Pe}$ and $\mathrm{Hy} = 2(1-\phi_0)/(\phi_m -\phi_0)$. $\mathrm{Hy}$ gives the ratio of the compressibility contribution over the hydrodynamic contribution. $\mathrm{Hy} = 2$ corresponds to the case where the two contributions cancel one another and the profile becomes perfectly flat in the condensed region. When $\mathrm{Hy}$  goes to infinity, the solution without hydrodynamic interactions is recovered. 

\begin{figure}
     \includegraphics[width=.85\columnwidth]{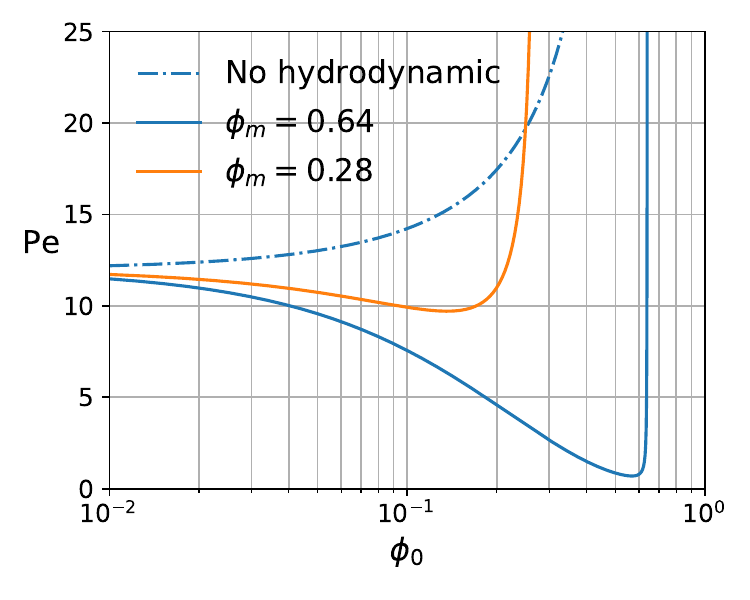}
    \caption{Peclet threshold for shell formation in presence of hydrodynamic interactions for different $\phi_m$.  }
\label{fig:Hydro_Pe}
\end{figure}

For the shell formation to occurs, the following condition $\xi<R$ on the characteristic length scale of the concentration profile has to be respected, leading to:
\begin{equation}
    \mathrm{Pe} > 12\frac{\phi_m(1-\phi_0)^6}{\phi_m - \phi_0}.
\end{equation}
This result highlights the effect of initial packing fraction $\phi_0$ on shell formation. We show Figure \ref{fig:Hydro_Pe} that the Peclet threshold for shell formation is non monotonic and presents a minimum value which shift towards $\phi_m$ as $\phi_m \rightarrow 1$. This indicates that for a given $\mathrm{Pe} < 12$, the shell forms lately in the droplet evaporation lifetime when the average packing fraction $\Bar{\phi} = \phi_0 (R_0 / R)^3$ increases and $\mathrm{Pe}$ becomes higher the threshold for this $\Bar{\phi}$.
As a side note, for $\phi_m = 0.64$ the random packing fraction for hard spheres, the packing fraction at which the threshold is minimum is equal to $0.57$ really close to $\phi_{\mathrm{g}} = 0.59$ the hard spheres glassy transition packing fraction.

\begin{figure}
     \includegraphics[width=.85\columnwidth]{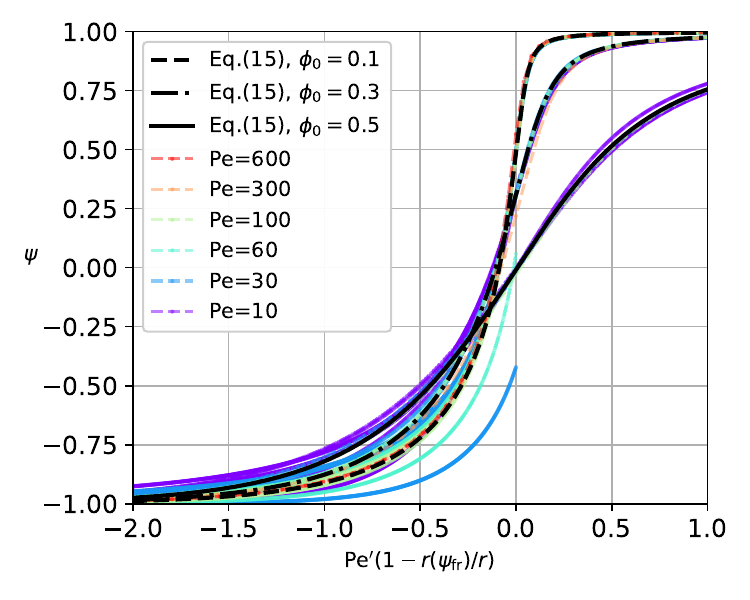}
    \caption{ Rescaled packing fraction profiles obtained from solutions \eqref{Eq:Sol_Hydro} as well as full numerical solution of \eqref{Eq:Gen_eq} for $\mathrm{Hy} = 0.1, 0.3$ and $0.5$ with $\phi_m = 0.64$. }
\label{fig:Hydro_prof}
\end{figure}

The effect of initial packing fraction is now considered on the concentration profile.
An implicit solution is obtained from equation \eqref{Eq:const_Hydro}, using simple element decomposition and polynomial long division:
\begin{multline}
    \frac{\mathrm{Pe'}}{6(1-\phi_0)^6} \frac{R}{R-h} \left( 1- \frac{R-h}{r} \right) = \\\left[ \arctanh{(\psi)} + \frac{\left(1-2/\mathrm{Hy}\right)^{6}}{1-\psi} \right.\\ +  \frac{1}{2}\left(1 - (1+10/\mathrm{Hy})(1-2/\mathrm{Hy})^5\right)\ln{(1-\psi)} \\ 
    + \frac{1}{2\mathrm{Hy}^6}\left(1+\psi\right)^4 -  \frac{4}{\mathrm{Hy}^5}\left(1-\frac{2}{3\mathrm{Hy}}\right)\left(1+\psi\right)^3 \\ + \frac{15}{\mathrm{Hy}^4}\left( 1- \frac{8}{5\mathrm{Hy}} + \frac{4}{5\mathrm{Hy}^2}\right)\left(1+\psi\right)^2 \\ \left. - \frac{40}{\mathrm{Hy}^3}\left(1-\frac{3}{\mathrm{Hy}}+\frac{18}{5\mathrm{Hy}^2} -\frac{8}{5\mathrm{Hy}^3}\right)\left(1+\psi\right) \right]^{\psi}_{\psi_{\mathrm{fr}}}.
    \label{Eq:Sol_Hydro}
\end{multline}
The solution still includes a hyperbolic arctangent term forming a front profile but now also includes power law terms with positive exponents smoothing the solution in the diluted region (see Figure \ref{fig:Hydro_prof}). The packing fraction at the front $\psi_{\mathrm{fr}}$ is given by:
\begin{equation}
    \psi_{\mathrm{fr}} = \frac{1}{2}\left[ \frac{4}{3} - \mathrm{Hy} +\sqrt{(\mathrm{Hy}-2)^2 + 7+\frac{1}{9} }  \right]
\end{equation}
converging asymptotically to $\psi_{\mathrm{fr}} \rightarrow -1/3$ when $\mathrm{Hy} \rightarrow \infty $ corresponding to the situation without hydrodynamic interaction and crossing $\psi_{\mathrm{fr}} = 0$ for $\mathrm{Hy} = 7$. 
The effect of $\mathrm{Hy}$ on the concentration profile is shown Figure \ref{fig:Hydro_prof} where the solution \eqref{Eq:Sol_Hydro} is compared to the finite-element numerical solution.

\begin{figure}
     \includegraphics[width=.85\columnwidth]{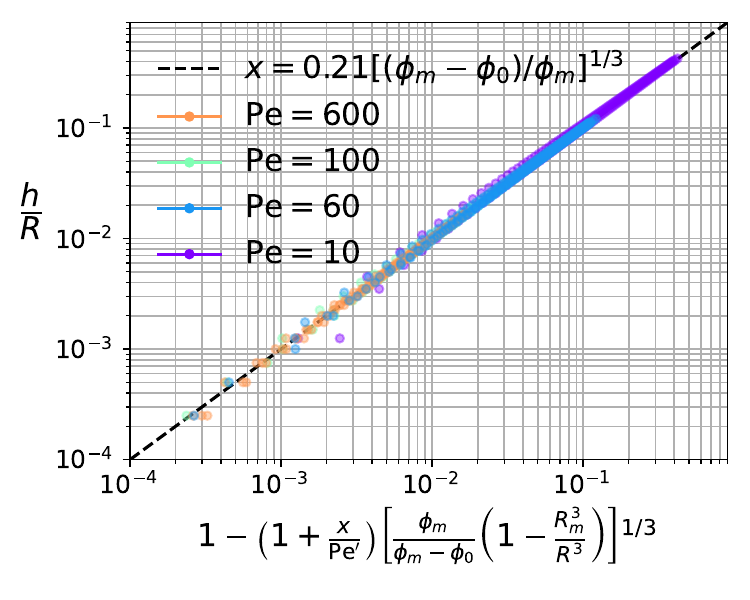}
    \caption{ Rescaled numerical shell thickness given from the front position compared to equation \eqref{Eq:Hydro_shell}. Here $\phi_0 = 0.1, 0.3, 0.5$ and $\phi_m = 0.64$. }
\label{fig:Hydro_shell}
\end{figure}

Surprisingly, compared to the case without hydrodynamic interactions, the concentration profile on the condensed region is more flat and relaxes more slowly. It also tends towards more symmetric solutions as $\phi_0$ increases. Since the condensed phase concentration profile is particular flat, equation \eqref{Eq:h} approximates well the front position up to a building time delay for the front to appear. In a volume conservation approach, this leads to shift the radius at which $h=0$. As this radius shift should be given by $\xi = R/\mathrm{Pe}'$ the shell intrinsic length scale, this leads to: 
\begin{equation}
    \frac{R-h}{R+x\xi} = \left( \frac{\phi_m}{\phi_m - \phi_0} \left[1- \left(\frac{R_{min}}{R}\right)^3\right] \right)^{1/3} 
    \label{Eq:Hydro_shell}
\end{equation}
where $x = 0.21\left(\frac{\phi_m - \phi_0}{\phi_m} \right)^{3}$ is fitted on the full numerical solutions (see Figure \ref{fig:Hydro_shell}). Here we still find correctly $h=R$ for $R=R_{min}$ and $h=0$ for $R<R_0$.  

Using this correction on the front position, this closes the analytical model and makes it almost equivalent to its numerical counterpart while being analytically tractable.

\section{Conclusion and discussion}\label{sec:conclusion}

We proposed here an analytical approach of general particles transport problem in confinement and applying to colloidal skin formation in evaporating droplet. This approach consists in finding a moving reference frame inspired by the confining domain boundary conditions where the problem can be considered as quasi-static.  On this moving reference frame, a constitutive equation can be explicitly derived leaving mainly the concentration at the boundary as unknown to solve the problem. 
For radially symmetric problems and in the shell formation regime, the boundary concentration can relied to the shell front position and an implicit solution can be found for any rational equation of state and mobility terms. This implicit solution has been explicitly derived for different compressibility terms only defined by their diverging exponent and in presence of hydrodynamic interactions, and is compared successfully to the high Peclet limit of full numerical solutions obtained using a finite-element method.  \\

To our knowledge, this is the first complete solution for this problem analytically tractable. Interestingly, it always shows a clear front resulting directly from the shrinking dynamics and is associated with a length scale $\xi \sim D_0 / \dot{R}$ with $R$ the domain radius and $D_0$ the self-diffusion coefficient. This front then becomes observable only for $R /\xi > 1 $ retrieving correctly the standard Peclet number role on the shell formation. Incidentally, it also shows that on the high Peclet limit, the shell front profile dynamically becomes sharper as the droplet evaporates. 
A 2D Brownian particles based simulation is also conducted as an ideal test of the continuum approach built only from first principles. This simulation is found to compare successfully qualitatively and quantitatively to the continuum to the continuum model up to its numerical limits. Particularly the shell front onset and the shell front intrinsic length scale is found to correspond to the one predicted, showing the robustness of this approach. 

Interestingly, the obtained concentration profile also shows an asymmetric profile observed in the particles based simulation and retrieved by the implicit solution, resulting from the difference between the condensed and diluted phase compressibility behavior. Including hydrodynamic interactions here can play a huge role, as their contribution to the particles interaction only depend on the void fraction and not on the maximum packing fraction.
 
This leads surprisingly to opposite asymmetric profile compared to the situation hydrodynamic interactions. In particular, the condensed phase concentration profile becomes markedly flat, thereby enabling the derivation of a novel empirical expression for the shell thickness that accounts for the condensed phase initial building time.
Varying the initial packing fraction, can modify significantly the concentration profile as the hydrodynamic influence becomes negligeable when $\phi_m - \phi_0 \ll 1 - \phi_0$ and leads to a non-monotonic behavior of the Peclet onset for shell formation where a minimum is found close to the maximum packing fraction. This means that delayed shell formation scenarii becomes accessible in a given range of Peclet number by using the initial packing fraction to acts simultaneously on the shell building time and on the shell formation Peclet threshold. \\

This delayed shell formation scenario should have a significant influence on the ultimate shell buckling as it affects the strain and Darcy’s pressure on the shell by varying the instantaneous ratio of the shell thickness over droplet radius, as well as the integrated amount of deformation sustained by the shell during the droplet evaporation life time. 
While the exact mechanism at the buckling origin has been subject to numerous discussions, a role is commonly attributed to the glassy transition characterized by its diverging relaxation time in the shell formation and the buckling onset \cite{tsapis2005onset, dauchot2023glass, lintingre2016control, milani2023double}.
While the first point about a local balance being the mechanism at the buckling origin has been subject to numerous discussions, we propose here that the history of the shell deformation and its accumulated mechanical constraints can also be at the origin of the buckling mechanism. 
This hypothesis is also reinforced by the role commonly attributed to the glassy transition characterized by its diverging relaxation time in the shell formation and the buckling onset \cite{tsapis2005onset, dauchot2023glass, lintingre2016control, milani2023double}.

\begin{figure}
     \includegraphics[width=.85\columnwidth]{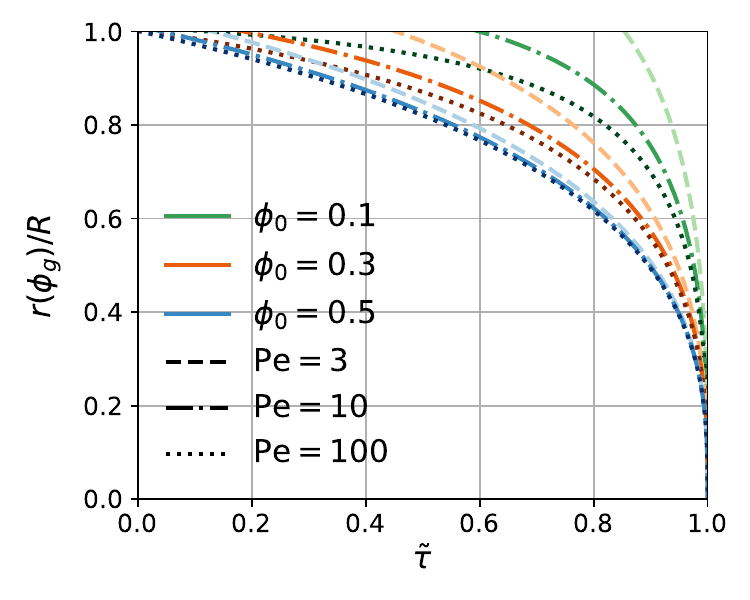}
    \caption{ Rescaled position where $\phi = \phi_{\mathrm{g}}$ with $\phi_{\mathrm{g}} = 0.59$ in function of rescaled time $\tilde{t}= t / \tau_{\mathrm{min}}$ with $\tau_{\mathrm{min}} = (R^2_0-R^2_{min}) / 2 \mathrm{Pe} D_0$. }
\label{fig:Hydro_glasspos}
\end{figure}

Using the concentration profile from Eq. \ref{Eq:Sol_Hydro} and the empiric shell thickness expression, we show in Fig. \ref{fig:Hydro_glasspos} the position of the glassy layer $r(\phi_{\mathrm{g}})$ in the evaporating droplet for different P\'eclet numbers and initial packing fractions. For high initial packing fractions, the P\'eclet number has negligible effect on the glassy layer position while for low initial packing, we observe a clear delayed shell formation.

\begin{figure}
     \includegraphics[width=.85\columnwidth]{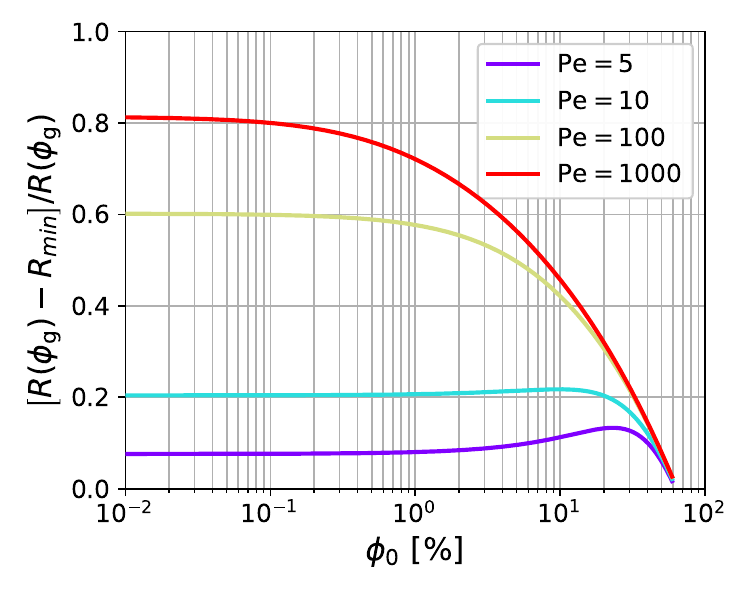}
    \caption{Relative maximum droplet retraction after the formation onset of a glassy colloidal phase.}
\label{fig:Hydro_DeltaRgel}
\end{figure}

With the data at hand, we can determine how much deformation does the glassy layer undergo by computing the difference between the radius at which the glassy layer begins to form and the minimum achievable radius $R(\phi_g)-R_{min}$, which can also give an estimate of the maximum expected aspect ratio of the resulting cluster. Figure \ref{fig:Hydro_DeltaRgel} shows this magnitude, normalized by $R(\phi_g)$, as a function of the initial packing fraction. 
The results shown in Fig. \ref{fig:Hydro_DeltaRgel} yield a good qualitative comparison with the experimental results of Seyfert et al. \cite{seyfert2021evaporation}. As shown in their experiments, low evaporation rates and low initial packing fractions yield less deformed supraballs. Here, the least-deformed structures are obtained when the initial packing fraction tends to the glassy transition packing fraction, predicting that the cluster geometry could also be altered by manipulating colloids interactions \cite{lintingre2016control}. 
Surprisingly, we observe a non-monotonic behavior with a maximum deformation which depends strongly on the P\'eclet number. 
This implies that buckling in such systems could be manipulated by using the right combination of humidity level and initial packing fraction.

However, supposing intrinsically that the concentration field is at local thermodynamic equilibrium, it can only leave open questions on the long time scales of phase transitions which are of utmost importance for understanding the mechanical properties of the glassy phase.
Still, this general approach allows for extension for more complex geometries or by adding other physical mechanisms. 
For example, to capture more satisfactorily the buckling mechanism, one should take into account the non-homogeneous nature of the evaporative flux, due to the presence of the substrate contact \cite{gelderblom2011water,wilson2023evaporation}.
Nonetheless, we think that this general approach could be useful to address a number of problems related to skin formation and more generally non-linear transport problem in confinement.

\begin{acknowledgements}
   AM and RS acknowledge funding from Horizon program of the European Research Council, proof of concept grant number 101101022. The authors also thank Eva Krolis for performing preliminary simulations that lead to this work.
   RS thanks Duarte Rocha for his help with the finite element solver.
\end{acknowledgements}


\begin{thebibliography}{39}%
\makeatletter
\providecommand \@ifxundefined [1]{%
 \@ifx{#1\undefined}
}%
\providecommand \@ifnum [1]{%
 \ifnum #1\expandafter \@firstoftwo
 \else \expandafter \@secondoftwo
 \fi
}%
\providecommand \@ifx [1]{%
 \ifx #1\expandafter \@firstoftwo
 \else \expandafter \@secondoftwo
 \fi
}%
\providecommand \natexlab [1]{#1}%
\providecommand \enquote  [1]{``#1''}%
\providecommand \bibnamefont  [1]{#1}%
\providecommand \bibfnamefont [1]{#1}%
\providecommand \citenamefont [1]{#1}%
\providecommand \href@noop [0]{\@secondoftwo}%
\providecommand \href [0]{\begingroup \@sanitize@url \@href}%
\providecommand \@href[1]{\@@startlink{#1}\@@href}%
\providecommand \@@href[1]{\endgroup#1\@@endlink}%
\providecommand \@sanitize@url [0]{\catcode `\\12\catcode `\$12\catcode
  `\&12\catcode `\#12\catcode `\^12\catcode `\_12\catcode `\%12\relax}%
\providecommand \@@startlink[1]{}%
\providecommand \@@endlink[0]{}%
\providecommand \url  [0]{\begingroup\@sanitize@url \@url }%
\providecommand \@url [1]{\endgroup\@href {#1}{\urlprefix }}%
\providecommand \urlprefix  [0]{URL }%
\providecommand \Eprint [0]{\href }%
\providecommand \doibase [0]{https://doi.org/}%
\providecommand \selectlanguage [0]{\@gobble}%
\providecommand \bibinfo  [0]{\@secondoftwo}%
\providecommand \bibfield  [0]{\@secondoftwo}%
\providecommand \translation [1]{[#1]}%
\providecommand \BibitemOpen [0]{}%
\providecommand \bibitemStop [0]{}%
\providecommand \bibitemNoStop [0]{.\EOS\space}%
\providecommand \EOS [0]{\spacefactor3000\relax}%
\providecommand \BibitemShut  [1]{\csname bibitem#1\endcsname}%
\let\auto@bib@innerbib\@empty
\bibitem [{\citenamefont {Deegan}\ \emph {et~al.}(1997)\citenamefont {Deegan},
  \citenamefont {Bakajin}, \citenamefont {Dupont}, \citenamefont {Huber},
  \citenamefont {Nagel},\ and\ \citenamefont {Witten}}]{deegan1997capillary}%
  \BibitemOpen
  \bibfield  {author} {\bibinfo {author} {\bibfnamefont {R.~D.}\ \bibnamefont
  {Deegan}}, \bibinfo {author} {\bibfnamefont {O.}~\bibnamefont {Bakajin}},
  \bibinfo {author} {\bibfnamefont {T.~F.}\ \bibnamefont {Dupont}}, \bibinfo
  {author} {\bibfnamefont {G.}~\bibnamefont {Huber}}, \bibinfo {author}
  {\bibfnamefont {S.~R.}\ \bibnamefont {Nagel}},\ and\ \bibinfo {author}
  {\bibfnamefont {T.~A.}\ \bibnamefont {Witten}},\ }\bibfield  {title}
  {\bibinfo {title} {Capillary flow as the cause of ring stains from dried
  liquid drops},\ }\href@noop {} {\bibfield  {journal} {\bibinfo  {journal}
  {Nature}\ }\textbf {\bibinfo {volume} {389}},\ \bibinfo {pages} {827}
  (\bibinfo {year} {1997})}\BibitemShut {NoStop}%
\bibitem [{\citenamefont {Gelderblom}\ \emph {et~al.}(2022)\citenamefont
  {Gelderblom}, \citenamefont {Diddens},\ and\ \citenamefont
  {Marin}}]{gelderblom2022evaporation}%
  \BibitemOpen
  \bibfield  {author} {\bibinfo {author} {\bibfnamefont {H.}~\bibnamefont
  {Gelderblom}}, \bibinfo {author} {\bibfnamefont {C.}~\bibnamefont
  {Diddens}},\ and\ \bibinfo {author} {\bibfnamefont {A.}~\bibnamefont
  {Marin}},\ }\bibfield  {title} {\bibinfo {title} {Evaporation-driven liquid
  flow in sessile droplets},\ }\href@noop {} {\bibfield  {journal} {\bibinfo
  {journal} {Soft Matter}\ }\textbf {\bibinfo {volume} {18}},\ \bibinfo {pages}
  {8535} (\bibinfo {year} {2022})}\BibitemShut {NoStop}%
\bibitem [{\citenamefont {Mar{\'\i}n}\ \emph {et~al.}(2012)\citenamefont
  {Mar{\'\i}n}, \citenamefont {Gelderblom}, \citenamefont {Susarrey-Arce},
  \citenamefont {van Houselt}, \citenamefont {Lefferts}, \citenamefont
  {Gardeniers}, \citenamefont {Lohse},\ and\ \citenamefont
  {Snoeijer}}]{marin2012building}%
  \BibitemOpen
  \bibfield  {author} {\bibinfo {author} {\bibfnamefont {{\'A}.~G.}\
  \bibnamefont {Mar{\'\i}n}}, \bibinfo {author} {\bibfnamefont
  {H.}~\bibnamefont {Gelderblom}}, \bibinfo {author} {\bibfnamefont
  {A.}~\bibnamefont {Susarrey-Arce}}, \bibinfo {author} {\bibfnamefont
  {A.}~\bibnamefont {van Houselt}}, \bibinfo {author} {\bibfnamefont
  {L.}~\bibnamefont {Lefferts}}, \bibinfo {author} {\bibfnamefont {J.~G.}\
  \bibnamefont {Gardeniers}}, \bibinfo {author} {\bibfnamefont
  {D.}~\bibnamefont {Lohse}},\ and\ \bibinfo {author} {\bibfnamefont {J.~H.}\
  \bibnamefont {Snoeijer}},\ }\bibfield  {title} {\bibinfo {title} {Building
  microscopic soccer balls with evaporating colloidal fakir drops},\
  }\href@noop {} {\bibfield  {journal} {\bibinfo  {journal} {Proceedings of the
  National Academy of Sciences}\ }\textbf {\bibinfo {volume} {109}},\ \bibinfo
  {pages} {16455} (\bibinfo {year} {2012})}\BibitemShut {NoStop}%
\bibitem [{\citenamefont {Wooh}\ \emph {et~al.}(2015)\citenamefont {Wooh},
  \citenamefont {Huesmann}, \citenamefont {Tahir}, \citenamefont {Paven},
  \citenamefont {Wichmann}, \citenamefont {Vollmer}, \citenamefont {Tremel},
  \citenamefont {Papadopoulos},\ and\ \citenamefont
  {Butt}}]{wooh2015synthesis}%
  \BibitemOpen
  \bibfield  {author} {\bibinfo {author} {\bibfnamefont {S.}~\bibnamefont
  {Wooh}}, \bibinfo {author} {\bibfnamefont {H.}~\bibnamefont {Huesmann}},
  \bibinfo {author} {\bibfnamefont {M.~N.}\ \bibnamefont {Tahir}}, \bibinfo
  {author} {\bibfnamefont {M.}~\bibnamefont {Paven}}, \bibinfo {author}
  {\bibfnamefont {K.}~\bibnamefont {Wichmann}}, \bibinfo {author}
  {\bibfnamefont {D.}~\bibnamefont {Vollmer}}, \bibinfo {author} {\bibfnamefont
  {W.}~\bibnamefont {Tremel}}, \bibinfo {author} {\bibfnamefont
  {P.}~\bibnamefont {Papadopoulos}},\ and\ \bibinfo {author} {\bibfnamefont
  {H.-J.}\ \bibnamefont {Butt}},\ }\bibfield  {title} {\bibinfo {title}
  {Synthesis of mesoporous supraparticles on superamphiphobic surfaces},\
  }\href@noop {} {\bibfield  {journal} {\bibinfo  {journal} {Advanced
  Materials}\ }\textbf {\bibinfo {volume} {27}},\ \bibinfo {pages} {7338}
  (\bibinfo {year} {2015})}\BibitemShut {NoStop}%
\bibitem [{\citenamefont {Seyfert}\ \emph {et~al.}(2021)\citenamefont
  {Seyfert}, \citenamefont {Berenschot}, \citenamefont {Tas}, \citenamefont
  {Susarrey-Arce},\ and\ \citenamefont {Marin}}]{seyfert2021evaporation}%
  \BibitemOpen
  \bibfield  {author} {\bibinfo {author} {\bibfnamefont {C.}~\bibnamefont
  {Seyfert}}, \bibinfo {author} {\bibfnamefont {E.~J.}\ \bibnamefont
  {Berenschot}}, \bibinfo {author} {\bibfnamefont {N.~R.}\ \bibnamefont {Tas}},
  \bibinfo {author} {\bibfnamefont {A.}~\bibnamefont {Susarrey-Arce}},\ and\
  \bibinfo {author} {\bibfnamefont {A.}~\bibnamefont {Marin}},\ }\bibfield
  {title} {\bibinfo {title} {Evaporation-driven colloidal cluster assembly
  using droplets on superhydrophobic fractal-like structures},\ }\href@noop {}
  {\bibfield  {journal} {\bibinfo  {journal} {Soft Matter}\ }\textbf {\bibinfo
  {volume} {17}},\ \bibinfo {pages} {506} (\bibinfo {year} {2021})}\BibitemShut
  {NoStop}%
\bibitem [{\citenamefont {Wang}\ \emph {et~al.}(2024)\citenamefont {Wang},
  \citenamefont {Lian}, \citenamefont {Xiang}, \citenamefont {Tao},
  \citenamefont {Kappl},\ and\ \citenamefont {Liu}}]{Wang2024Supraparticle}%
  \BibitemOpen
  \bibfield  {author} {\bibinfo {author} {\bibfnamefont {X.}~\bibnamefont
  {Wang}}, \bibinfo {author} {\bibfnamefont {Y.}~\bibnamefont {Lian}}, \bibinfo
  {author} {\bibfnamefont {S.}~\bibnamefont {Xiang}}, \bibinfo {author}
  {\bibfnamefont {S.}~\bibnamefont {Tao}}, \bibinfo {author} {\bibfnamefont
  {M.}~\bibnamefont {Kappl}},\ and\ \bibinfo {author} {\bibfnamefont
  {W.}~\bibnamefont {Liu}},\ }\bibfield  {title} {\bibinfo {title} {Droplet
  evaporation on super liquid-repellent surfaces: A controllable approach for
  supraparticle fabrication},\ }\href@noop {} {\bibfield  {journal} {\bibinfo
  {journal} {Advances in Colloid and Interface Science}\ ,\ \bibinfo {pages}
  {103305}} (\bibinfo {year} {2024})}\BibitemShut {NoStop}%
\bibitem [{\citenamefont {Wintzheimer}\ \emph {et~al.}(2021)\citenamefont
  {Wintzheimer}, \citenamefont {Reichstein}, \citenamefont {Groppe},
  \citenamefont {Wolf}, \citenamefont {Fett}, \citenamefont {Zhou},
  \citenamefont {Pujales-Paradela}, \citenamefont {Miller}, \citenamefont
  {M{\"u}ssig}, \citenamefont {Wenderoth} \emph
  {et~al.}}]{wintzheimer2021supraparticles}%
  \BibitemOpen
  \bibfield  {author} {\bibinfo {author} {\bibfnamefont {S.}~\bibnamefont
  {Wintzheimer}}, \bibinfo {author} {\bibfnamefont {J.}~\bibnamefont
  {Reichstein}}, \bibinfo {author} {\bibfnamefont {P.}~\bibnamefont {Groppe}},
  \bibinfo {author} {\bibfnamefont {A.}~\bibnamefont {Wolf}}, \bibinfo {author}
  {\bibfnamefont {B.}~\bibnamefont {Fett}}, \bibinfo {author} {\bibfnamefont
  {H.}~\bibnamefont {Zhou}}, \bibinfo {author} {\bibfnamefont {R.}~\bibnamefont
  {Pujales-Paradela}}, \bibinfo {author} {\bibfnamefont {F.}~\bibnamefont
  {Miller}}, \bibinfo {author} {\bibfnamefont {S.}~\bibnamefont {M{\"u}ssig}},
  \bibinfo {author} {\bibfnamefont {S.}~\bibnamefont {Wenderoth}}, \emph
  {et~al.},\ }\bibfield  {title} {\bibinfo {title} {Supraparticles for
  sustainability},\ }\href@noop {} {\bibfield  {journal} {\bibinfo  {journal}
  {Advanced Functional Materials}\ }\textbf {\bibinfo {volume} {31}},\ \bibinfo
  {pages} {2011089} (\bibinfo {year} {2021})}\BibitemShut {NoStop}%
\bibitem [{\citenamefont {Basu}\ \emph {et~al.}(2016)\citenamefont {Basu},
  \citenamefont {Bansal},\ and\ \citenamefont {Miglani}}]{basu2016towards}%
  \BibitemOpen
  \bibfield  {author} {\bibinfo {author} {\bibfnamefont {S.}~\bibnamefont
  {Basu}}, \bibinfo {author} {\bibfnamefont {L.}~\bibnamefont {Bansal}},\ and\
  \bibinfo {author} {\bibfnamefont {A.}~\bibnamefont {Miglani}},\ }\bibfield
  {title} {\bibinfo {title} {Towards universal buckling dynamics in
  nanocolloidal sessile droplets: the effect of hydrophilic to superhydrophobic
  substrates and evaporation modes},\ }\href@noop {} {\bibfield  {journal}
  {\bibinfo  {journal} {Soft matter}\ }\textbf {\bibinfo {volume} {12}},\
  \bibinfo {pages} {4896} (\bibinfo {year} {2016})}\BibitemShut {NoStop}%
\bibitem [{\citenamefont {Tsapis}\ \emph {et~al.}(2005)\citenamefont {Tsapis},
  \citenamefont {Dufresne}, \citenamefont {Sinha}, \citenamefont {Riera},
  \citenamefont {Hutchinson}, \citenamefont {Mahadevan},\ and\ \citenamefont
  {Weitz}}]{tsapis2005onset}%
  \BibitemOpen
  \bibfield  {author} {\bibinfo {author} {\bibfnamefont {N.}~\bibnamefont
  {Tsapis}}, \bibinfo {author} {\bibfnamefont {E.~R.}\ \bibnamefont
  {Dufresne}}, \bibinfo {author} {\bibfnamefont {S.~S.}\ \bibnamefont {Sinha}},
  \bibinfo {author} {\bibfnamefont {C.~S.}\ \bibnamefont {Riera}}, \bibinfo
  {author} {\bibfnamefont {J.~W.}\ \bibnamefont {Hutchinson}}, \bibinfo
  {author} {\bibfnamefont {L.}~\bibnamefont {Mahadevan}},\ and\ \bibinfo
  {author} {\bibfnamefont {D.~A.}\ \bibnamefont {Weitz}},\ }\bibfield  {title}
  {\bibinfo {title} {Onset of buckling in drying droplets of colloidal
  suspensions},\ }\href@noop {} {\bibfield  {journal} {\bibinfo  {journal}
  {Physical review letters}\ }\textbf {\bibinfo {volume} {94}},\ \bibinfo
  {pages} {018302} (\bibinfo {year} {2005})}\BibitemShut {NoStop}%
\bibitem [{\citenamefont {Archer}\ \emph {et~al.}(2020)\citenamefont {Archer},
  \citenamefont {Walker}, \citenamefont {Gregson}, \citenamefont {Hardy},\ and\
  \citenamefont {Reid}}]{archer2020drying}%
  \BibitemOpen
  \bibfield  {author} {\bibinfo {author} {\bibfnamefont {J.}~\bibnamefont
  {Archer}}, \bibinfo {author} {\bibfnamefont {J.~S.}\ \bibnamefont {Walker}},
  \bibinfo {author} {\bibfnamefont {F.~K.}\ \bibnamefont {Gregson}}, \bibinfo
  {author} {\bibfnamefont {D.~A.}\ \bibnamefont {Hardy}},\ and\ \bibinfo
  {author} {\bibfnamefont {J.~P.}\ \bibnamefont {Reid}},\ }\bibfield  {title}
  {\bibinfo {title} {Drying kinetics and particle formation from dilute
  colloidal suspensions in aerosol droplets},\ }\href@noop {} {\bibfield
  {journal} {\bibinfo  {journal} {Langmuir}\ }\textbf {\bibinfo {volume}
  {36}},\ \bibinfo {pages} {12481} (\bibinfo {year} {2020})}\BibitemShut
  {NoStop}%
\bibitem [{\citenamefont {Lintingre}\ \emph {et~al.}(2016)\citenamefont
  {Lintingre}, \citenamefont {Lequeux}, \citenamefont {Talini},\ and\
  \citenamefont {Tsapis}}]{lintingre2016control}%
  \BibitemOpen
  \bibfield  {author} {\bibinfo {author} {\bibfnamefont {E.}~\bibnamefont
  {Lintingre}}, \bibinfo {author} {\bibfnamefont {F.}~\bibnamefont {Lequeux}},
  \bibinfo {author} {\bibfnamefont {L.}~\bibnamefont {Talini}},\ and\ \bibinfo
  {author} {\bibfnamefont {N.}~\bibnamefont {Tsapis}},\ }\bibfield  {title}
  {\bibinfo {title} {Control of particle morphology in the spray drying of
  colloidal suspensions},\ }\href@noop {} {\bibfield  {journal} {\bibinfo
  {journal} {Soft Matter}\ }\textbf {\bibinfo {volume} {12}},\ \bibinfo {pages}
  {7435} (\bibinfo {year} {2016})}\BibitemShut {NoStop}%
\bibitem [{\citenamefont {Boulogne}\ \emph {et~al.}(2013)\citenamefont
  {Boulogne}, \citenamefont {Giorgiutti-Dauphin{\'e}},\ and\ \citenamefont
  {Pauchard}}]{boulogne2013buckling}%
  \BibitemOpen
  \bibfield  {author} {\bibinfo {author} {\bibfnamefont {F.}~\bibnamefont
  {Boulogne}}, \bibinfo {author} {\bibfnamefont {F.}~\bibnamefont
  {Giorgiutti-Dauphin{\'e}}},\ and\ \bibinfo {author} {\bibfnamefont
  {L.}~\bibnamefont {Pauchard}},\ }\bibfield  {title} {\bibinfo {title} {The
  buckling and invagination process during consolidation of colloidal
  droplets},\ }\href@noop {} {\bibfield  {journal} {\bibinfo  {journal} {Soft
  Matter}\ }\textbf {\bibinfo {volume} {9}},\ \bibinfo {pages} {750} (\bibinfo
  {year} {2013})}\BibitemShut {NoStop}%
\bibitem [{\citenamefont {Bouchaudy}\ and\ \citenamefont
  {Salmon}(2019)}]{bouchaudy2019drying}%
  \BibitemOpen
  \bibfield  {author} {\bibinfo {author} {\bibfnamefont {A.}~\bibnamefont
  {Bouchaudy}}\ and\ \bibinfo {author} {\bibfnamefont {J.-B.}\ \bibnamefont
  {Salmon}},\ }\bibfield  {title} {\bibinfo {title} {Drying-induced stresses
  before solidification in colloidal dispersions: In situ measurements},\
  }\href@noop {} {\bibfield  {journal} {\bibinfo  {journal} {Soft Matter}\
  }\textbf {\bibinfo {volume} {15}},\ \bibinfo {pages} {2768} (\bibinfo {year}
  {2019})}\BibitemShut {NoStop}%
\bibitem [{\citenamefont {Sobac}\ \emph {et~al.}(2019)\citenamefont {Sobac},
  \citenamefont {Larbi}, \citenamefont {Colinet},\ and\ \citenamefont
  {Haut}}]{sobac2019mathematical}%
  \BibitemOpen
  \bibfield  {author} {\bibinfo {author} {\bibfnamefont {B.}~\bibnamefont
  {Sobac}}, \bibinfo {author} {\bibfnamefont {Z.}~\bibnamefont {Larbi}},
  \bibinfo {author} {\bibfnamefont {P.}~\bibnamefont {Colinet}},\ and\ \bibinfo
  {author} {\bibfnamefont {B.}~\bibnamefont {Haut}},\ }\bibfield  {title}
  {\bibinfo {title} {Mathematical modeling of the drying of a spherical
  colloidal drop},\ }\href@noop {} {\bibfield  {journal} {\bibinfo  {journal}
  {Colloids and Surfaces A: Physicochemical and Engineering Aspects}\ }\textbf
  {\bibinfo {volume} {576}},\ \bibinfo {pages} {110} (\bibinfo {year}
  {2019})}\BibitemShut {NoStop}%
\bibitem [{\citenamefont {Daubersies}\ and\ \citenamefont
  {Salmon}(2011)}]{daubersies2011evaporation}%
  \BibitemOpen
  \bibfield  {author} {\bibinfo {author} {\bibfnamefont {L.}~\bibnamefont
  {Daubersies}}\ and\ \bibinfo {author} {\bibfnamefont {J.-B.}\ \bibnamefont
  {Salmon}},\ }\bibfield  {title} {\bibinfo {title} {Evaporation of solutions
  and colloidal dispersions in confined droplets},\ }\href@noop {} {\bibfield
  {journal} {\bibinfo  {journal} {Physical Review E}\ }\textbf {\bibinfo
  {volume} {84}},\ \bibinfo {pages} {031406} (\bibinfo {year}
  {2011})}\BibitemShut {NoStop}%
\bibitem [{\citenamefont {Loussert}\ \emph {et~al.}(2016)\citenamefont
  {Loussert}, \citenamefont {Bouchaudy},\ and\ \citenamefont
  {Salmon}}]{loussert2016drying}%
  \BibitemOpen
  \bibfield  {author} {\bibinfo {author} {\bibfnamefont {C.}~\bibnamefont
  {Loussert}}, \bibinfo {author} {\bibfnamefont {A.}~\bibnamefont
  {Bouchaudy}},\ and\ \bibinfo {author} {\bibfnamefont {J.-B.}\ \bibnamefont
  {Salmon}},\ }\bibfield  {title} {\bibinfo {title} {Drying dynamics of a
  charged colloidal dispersion in a confined drop},\ }\href@noop {} {\bibfield
  {journal} {\bibinfo  {journal} {Physical Review Fluids}\ }\textbf {\bibinfo
  {volume} {1}},\ \bibinfo {pages} {084201} (\bibinfo {year}
  {2016})}\BibitemShut {NoStop}%
\bibitem [{\citenamefont {Sobac}\ \emph {et~al.}(2020)\citenamefont {Sobac},
  \citenamefont {Dehaeck}, \citenamefont {Bouchaudy},\ and\ \citenamefont
  {Salmon}}]{sobac2020collective}%
  \BibitemOpen
  \bibfield  {author} {\bibinfo {author} {\bibfnamefont {B.}~\bibnamefont
  {Sobac}}, \bibinfo {author} {\bibfnamefont {S.}~\bibnamefont {Dehaeck}},
  \bibinfo {author} {\bibfnamefont {A.}~\bibnamefont {Bouchaudy}},\ and\
  \bibinfo {author} {\bibfnamefont {J.-B.}\ \bibnamefont {Salmon}},\ }\bibfield
   {title} {\bibinfo {title} {Collective diffusion coefficient of a charged
  colloidal dispersion: interferometric measurements in a drying drop},\
  }\href@noop {} {\bibfield  {journal} {\bibinfo  {journal} {Soft matter}\
  }\textbf {\bibinfo {volume} {16}},\ \bibinfo {pages} {8213} (\bibinfo {year}
  {2020})}\BibitemShut {NoStop}%
\bibitem [{\citenamefont {Drew}(1982)}]{drew1982mathematical}%
  \BibitemOpen
  \bibfield  {author} {\bibinfo {author} {\bibfnamefont {D.~A.}\ \bibnamefont
  {Drew}},\ }\bibfield  {title} {\bibinfo {title} {Mathematical modeling of
  two-phase flow},\ }\href@noop {} {\bibfield  {journal} {\bibinfo  {journal}
  {Technical Summary Report Wisconsin Univ}\ } (\bibinfo {year}
  {1982})}\BibitemShut {NoStop}%
\bibitem [{\citenamefont {Zhang}\ and\ \citenamefont
  {Prosperetti}(1997)}]{zhang1997momentum}%
  \BibitemOpen
  \bibfield  {author} {\bibinfo {author} {\bibfnamefont {D.}~\bibnamefont
  {Zhang}}\ and\ \bibinfo {author} {\bibfnamefont {A.}~\bibnamefont
  {Prosperetti}},\ }\bibfield  {title} {\bibinfo {title} {Momentum and energy
  equations for disperse two-phase flows and their closure for dilute
  suspensions},\ }\href@noop {} {\bibfield  {journal} {\bibinfo  {journal}
  {International journal of multiphase flow}\ }\textbf {\bibinfo {volume}
  {23}},\ \bibinfo {pages} {425} (\bibinfo {year} {1997})}\BibitemShut
  {NoStop}%
\bibitem [{\citenamefont {Nott}\ \emph {et~al.}(2011)\citenamefont {Nott},
  \citenamefont {Guazzelli},\ and\ \citenamefont
  {Pouliquen}}]{nott2011suspension}%
  \BibitemOpen
  \bibfield  {author} {\bibinfo {author} {\bibfnamefont {P.~R.}\ \bibnamefont
  {Nott}}, \bibinfo {author} {\bibfnamefont {E.}~\bibnamefont {Guazzelli}},\
  and\ \bibinfo {author} {\bibfnamefont {O.}~\bibnamefont {Pouliquen}},\
  }\bibfield  {title} {\bibinfo {title} {The suspension balance model
  revisited},\ }\href@noop {} {\bibfield  {journal} {\bibinfo  {journal}
  {Physics of Fluids}\ }\textbf {\bibinfo {volume} {23}} (\bibinfo {year}
  {2011})}\BibitemShut {NoStop}%
\bibitem [{\citenamefont {Squires}\ and\ \citenamefont
  {Mason}(2010)}]{squires2010fluid}%
  \BibitemOpen
  \bibfield  {author} {\bibinfo {author} {\bibfnamefont {T.~M.}\ \bibnamefont
  {Squires}}\ and\ \bibinfo {author} {\bibfnamefont {T.~G.}\ \bibnamefont
  {Mason}},\ }\bibfield  {title} {\bibinfo {title} {Fluid mechanics of
  microrheology},\ }\href@noop {} {\bibfield  {journal} {\bibinfo  {journal}
  {Annual review of fluid mechanics}\ }\textbf {\bibinfo {volume} {42}},\
  \bibinfo {pages} {413} (\bibinfo {year} {2010})}\BibitemShut {NoStop}%
\bibitem [{\citenamefont {Deboeuf}\ \emph {et~al.}(2009)\citenamefont
  {Deboeuf}, \citenamefont {Gauthier}, \citenamefont {Martin}, \citenamefont
  {Yurkovetsky},\ and\ \citenamefont {Morris}}]{deboeuf2009particle}%
  \BibitemOpen
  \bibfield  {author} {\bibinfo {author} {\bibfnamefont {A.}~\bibnamefont
  {Deboeuf}}, \bibinfo {author} {\bibfnamefont {G.}~\bibnamefont {Gauthier}},
  \bibinfo {author} {\bibfnamefont {J.}~\bibnamefont {Martin}}, \bibinfo
  {author} {\bibfnamefont {Y.}~\bibnamefont {Yurkovetsky}},\ and\ \bibinfo
  {author} {\bibfnamefont {J.~F.}\ \bibnamefont {Morris}},\ }\bibfield  {title}
  {\bibinfo {title} {Particle pressure in a sheared suspension: A bridge from
  osmosis to granular dilatancy},\ }\href@noop {} {\bibfield  {journal}
  {\bibinfo  {journal} {Physical review letters}\ }\textbf {\bibinfo {volume}
  {102}},\ \bibinfo {pages} {108301} (\bibinfo {year} {2009})}\BibitemShut
  {NoStop}%
\bibitem [{\citenamefont {Diddens}(2017)}]{diddens2017detailed}%
  \BibitemOpen
  \bibfield  {author} {\bibinfo {author} {\bibfnamefont {C.}~\bibnamefont
  {Diddens}},\ }\bibfield  {title} {\bibinfo {title} {Detailed finite element
  method modeling of evaporating multi-component droplets},\ }\href@noop {}
  {\bibfield  {journal} {\bibinfo  {journal} {Journal of Computational
  Physics}\ }\textbf {\bibinfo {volume} {340}},\ \bibinfo {pages} {670}
  (\bibinfo {year} {2017})}\BibitemShut {NoStop}%
\bibitem [{\citenamefont {Diddens}\ \emph {et~al.}(2017)\citenamefont
  {Diddens}, \citenamefont {Tan}, \citenamefont {Lv}, \citenamefont {Versluis},
  \citenamefont {Kuerten}, \citenamefont {Zhang},\ and\ \citenamefont
  {Lohse}}]{diddens2017evaporating}%
  \BibitemOpen
  \bibfield  {author} {\bibinfo {author} {\bibfnamefont {C.}~\bibnamefont
  {Diddens}}, \bibinfo {author} {\bibfnamefont {H.}~\bibnamefont {Tan}},
  \bibinfo {author} {\bibfnamefont {P.}~\bibnamefont {Lv}}, \bibinfo {author}
  {\bibfnamefont {M.}~\bibnamefont {Versluis}}, \bibinfo {author}
  {\bibfnamefont {J.}~\bibnamefont {Kuerten}}, \bibinfo {author} {\bibfnamefont
  {X.}~\bibnamefont {Zhang}},\ and\ \bibinfo {author} {\bibfnamefont
  {D.}~\bibnamefont {Lohse}},\ }\bibfield  {title} {\bibinfo {title}
  {Evaporating pure, binary and ternary droplets: thermal effects and axial
  symmetry breaking},\ }\href@noop {} {\bibfield  {journal} {\bibinfo
  {journal} {Journal of fluid mechanics}\ }\textbf {\bibinfo {volume} {823}},\
  \bibinfo {pages} {470} (\bibinfo {year} {2017})}\BibitemShut {NoStop}%
\bibitem [{\citenamefont {Routh}\ and\ \citenamefont
  {Zimmerman}(2004)}]{routh2004distribution}%
  \BibitemOpen
  \bibfield  {author} {\bibinfo {author} {\bibfnamefont {A.~F.}\ \bibnamefont
  {Routh}}\ and\ \bibinfo {author} {\bibfnamefont {W.~B.}\ \bibnamefont
  {Zimmerman}},\ }\bibfield  {title} {\bibinfo {title} {Distribution of
  particles during solvent evaporation from films},\ }\href@noop {} {\bibfield
  {journal} {\bibinfo  {journal} {Chemical Engineering Science}\ }\textbf
  {\bibinfo {volume} {59}},\ \bibinfo {pages} {2961} (\bibinfo {year}
  {2004})}\BibitemShut {NoStop}%
\bibitem [{\citenamefont {Langmuir}(1918)}]{Langmuir:1918}%
  \BibitemOpen
  \bibfield  {author} {\bibinfo {author} {\bibfnamefont {I.}~\bibnamefont
  {Langmuir}},\ }\bibfield  {title} {\bibinfo {title} {The evaporation of small
  spheres},\ }\href@noop {} {\bibfield  {journal} {\bibinfo  {journal} {Phys.
  Rev.}\ }\textbf {\bibinfo {volume} {12}},\  (\bibinfo {year}
  {1918})}\BibitemShut {NoStop}%
\bibitem [{\citenamefont {Milani}\ \emph {et~al.}(2023)\citenamefont {Milani},
  \citenamefont {Phou}, \citenamefont {Ligoure}, \citenamefont {Cipelletti},\
  and\ \citenamefont {Ramos}}]{milani2023double}%
  \BibitemOpen
  \bibfield  {author} {\bibinfo {author} {\bibfnamefont {M.}~\bibnamefont
  {Milani}}, \bibinfo {author} {\bibfnamefont {T.}~\bibnamefont {Phou}},
  \bibinfo {author} {\bibfnamefont {C.}~\bibnamefont {Ligoure}}, \bibinfo
  {author} {\bibfnamefont {L.}~\bibnamefont {Cipelletti}},\ and\ \bibinfo
  {author} {\bibfnamefont {L.}~\bibnamefont {Ramos}},\ }\bibfield  {title}
  {\bibinfo {title} {A double rigidity transition rules the fate of drying
  colloidal drops},\ }\href@noop {} {\bibfield  {journal} {\bibinfo  {journal}
  {Soft Matter}\ }\textbf {\bibinfo {volume} {19}},\ \bibinfo {pages} {6968}
  (\bibinfo {year} {2023})}\BibitemShut {NoStop}%
\bibitem [{\citenamefont {Ladiges}\ \emph {et~al.}(2021)\citenamefont
  {Ladiges}, \citenamefont {Nonaka}, \citenamefont {Klymko}, \citenamefont
  {Moore}, \citenamefont {Bell}, \citenamefont {Carney}, \citenamefont
  {Garcia}, \citenamefont {Natesh},\ and\ \citenamefont
  {Donev}}]{ladiges2021discrete}%
  \BibitemOpen
  \bibfield  {author} {\bibinfo {author} {\bibfnamefont {D.~R.}\ \bibnamefont
  {Ladiges}}, \bibinfo {author} {\bibfnamefont {A.}~\bibnamefont {Nonaka}},
  \bibinfo {author} {\bibfnamefont {K.}~\bibnamefont {Klymko}}, \bibinfo
  {author} {\bibfnamefont {G.}~\bibnamefont {Moore}}, \bibinfo {author}
  {\bibfnamefont {J.}~\bibnamefont {Bell}}, \bibinfo {author} {\bibfnamefont
  {S.}~\bibnamefont {Carney}}, \bibinfo {author} {\bibfnamefont
  {A.}~\bibnamefont {Garcia}}, \bibinfo {author} {\bibfnamefont
  {S.}~\bibnamefont {Natesh}},\ and\ \bibinfo {author} {\bibfnamefont
  {A.}~\bibnamefont {Donev}},\ }\bibfield  {title} {\bibinfo {title} {Discrete
  ion stochastic continuum overdamped solvent algorithm for modeling
  electrolytes},\ }\href@noop {} {\bibfield  {journal} {\bibinfo  {journal}
  {Physical Review Fluids}\ }\textbf {\bibinfo {volume} {6}},\ \bibinfo {pages}
  {044309} (\bibinfo {year} {2021})}\BibitemShut {NoStop}%
\bibitem [{\citenamefont {Buzzaccaro}\ \emph {et~al.}(2007)\citenamefont
  {Buzzaccaro}, \citenamefont {Rusconi},\ and\ \citenamefont
  {Piazza}}]{buzzaccaro2007sticky}%
  \BibitemOpen
  \bibfield  {author} {\bibinfo {author} {\bibfnamefont {S.}~\bibnamefont
  {Buzzaccaro}}, \bibinfo {author} {\bibfnamefont {R.}~\bibnamefont
  {Rusconi}},\ and\ \bibinfo {author} {\bibfnamefont {R.}~\bibnamefont
  {Piazza}},\ }\bibfield  {title} {\bibinfo {title} {“sticky” hard spheres:
  equation of state, phase diagram, and metastable gels},\ }\href@noop {}
  {\bibfield  {journal} {\bibinfo  {journal} {Physical review letters}\
  }\textbf {\bibinfo {volume} {99}},\ \bibinfo {pages} {098301} (\bibinfo
  {year} {2007})}\BibitemShut {NoStop}%
\bibitem [{\citenamefont {Piazza}(2014)}]{piazza2014settled}%
  \BibitemOpen
  \bibfield  {author} {\bibinfo {author} {\bibfnamefont {R.}~\bibnamefont
  {Piazza}},\ }\bibfield  {title} {\bibinfo {title} {Settled and unsettled
  issues in particle settling},\ }\href@noop {} {\bibfield  {journal} {\bibinfo
   {journal} {Reports on Progress in Physics}\ }\textbf {\bibinfo {volume}
  {77}},\ \bibinfo {pages} {056602} (\bibinfo {year} {2014})}\BibitemShut
  {NoStop}%
\bibitem [{\citenamefont {Daanoun}\ \emph {et~al.}(1994)\citenamefont
  {Daanoun}, \citenamefont {Tejero},\ and\ \citenamefont
  {Baus}}]{daanoun1994van}%
  \BibitemOpen
  \bibfield  {author} {\bibinfo {author} {\bibfnamefont {A.}~\bibnamefont
  {Daanoun}}, \bibinfo {author} {\bibfnamefont {C.}~\bibnamefont {Tejero}},\
  and\ \bibinfo {author} {\bibfnamefont {M.}~\bibnamefont {Baus}},\ }\bibfield
  {title} {\bibinfo {title} {van der waals theory for solids},\ }\href@noop {}
  {\bibfield  {journal} {\bibinfo  {journal} {Physical Review E}\ }\textbf
  {\bibinfo {volume} {50}},\ \bibinfo {pages} {2913} (\bibinfo {year}
  {1994})}\BibitemShut {NoStop}%
\bibitem [{\citenamefont {Richardson}\ and\ \citenamefont
  {Zaki}(1954)}]{richardson1954sedimentation}%
  \BibitemOpen
  \bibfield  {author} {\bibinfo {author} {\bibfnamefont {J.}~\bibnamefont
  {Richardson}}\ and\ \bibinfo {author} {\bibfnamefont {W.}~\bibnamefont
  {Zaki}},\ }\bibfield  {title} {\bibinfo {title} {The sedimentation of a
  suspension of uniform spheres under conditions of viscous flow},\ }\href@noop
  {} {\bibfield  {journal} {\bibinfo  {journal} {Chemical Engineering Science}\
  }\textbf {\bibinfo {volume} {3}},\ \bibinfo {pages} {65} (\bibinfo {year}
  {1954})}\BibitemShut {NoStop}%
\bibitem [{\citenamefont {Peppin}\ \emph {et~al.}(2006)\citenamefont {Peppin},
  \citenamefont {Elliott},\ and\ \citenamefont
  {Worster}}]{peppin2006solidification}%
  \BibitemOpen
  \bibfield  {author} {\bibinfo {author} {\bibfnamefont {S.}~\bibnamefont
  {Peppin}}, \bibinfo {author} {\bibfnamefont {J.}~\bibnamefont {Elliott}},\
  and\ \bibinfo {author} {\bibfnamefont {M.~G.}\ \bibnamefont {Worster}},\
  }\bibfield  {title} {\bibinfo {title} {Solidification of colloidal
  suspensions},\ }\href@noop {} {\bibfield  {journal} {\bibinfo  {journal}
  {Journal of Fluid Mechanics}\ }\textbf {\bibinfo {volume} {554}},\ \bibinfo
  {pages} {147} (\bibinfo {year} {2006})}\BibitemShut {NoStop}%
\bibitem [{\citenamefont {Russel}\ \emph {et~al.}(1991)\citenamefont {Russel},
  \citenamefont {Russel}, \citenamefont {Saville},\ and\ \citenamefont
  {Schowalter}}]{russel1991colloidal}%
  \BibitemOpen
  \bibfield  {author} {\bibinfo {author} {\bibfnamefont {W.~B.}\ \bibnamefont
  {Russel}}, \bibinfo {author} {\bibfnamefont {W.}~\bibnamefont {Russel}},
  \bibinfo {author} {\bibfnamefont {D.~A.}\ \bibnamefont {Saville}},\ and\
  \bibinfo {author} {\bibfnamefont {W.~R.}\ \bibnamefont {Schowalter}},\
  }\href@noop {} {\emph {\bibinfo {title} {Colloidal dispersions}}}\ (\bibinfo
  {publisher} {Cambridge university press},\ \bibinfo {year}
  {1991})\BibitemShut {NoStop}%
\bibitem [{\citenamefont {Dauchot}\ \emph {et~al.}(2023)\citenamefont
  {Dauchot}, \citenamefont {Ladieu},\ and\ \citenamefont
  {Royall}}]{dauchot2023glass}%
  \BibitemOpen
  \bibfield  {author} {\bibinfo {author} {\bibfnamefont {O.}~\bibnamefont
  {Dauchot}}, \bibinfo {author} {\bibfnamefont {F.}~\bibnamefont {Ladieu}},\
  and\ \bibinfo {author} {\bibfnamefont {C.~P.}\ \bibnamefont {Royall}},\
  }\bibfield  {title} {\bibinfo {title} {The glass transition in molecules,
  colloids and grains: universality and specificity},\ }\href@noop {}
  {\bibfield  {journal} {\bibinfo  {journal} {Comptes Rendus. Physique}\
  }\textbf {\bibinfo {volume} {24}},\ \bibinfo {pages} {1} (\bibinfo {year}
  {2023})}\BibitemShut {NoStop}%
\bibitem [{\citenamefont {Gelderblom}\ \emph {et~al.}(2011)\citenamefont
  {Gelderblom}, \citenamefont {Marin}, \citenamefont {Nair}, \citenamefont
  {Van~Houselt}, \citenamefont {Lefferts}, \citenamefont {Snoeijer},\ and\
  \citenamefont {Lohse}}]{gelderblom2011water}%
  \BibitemOpen
  \bibfield  {author} {\bibinfo {author} {\bibfnamefont {H.}~\bibnamefont
  {Gelderblom}}, \bibinfo {author} {\bibfnamefont {A.~G.}\ \bibnamefont
  {Marin}}, \bibinfo {author} {\bibfnamefont {H.}~\bibnamefont {Nair}},
  \bibinfo {author} {\bibfnamefont {A.}~\bibnamefont {Van~Houselt}}, \bibinfo
  {author} {\bibfnamefont {L.}~\bibnamefont {Lefferts}}, \bibinfo {author}
  {\bibfnamefont {J.~H.}\ \bibnamefont {Snoeijer}},\ and\ \bibinfo {author}
  {\bibfnamefont {D.}~\bibnamefont {Lohse}},\ }\bibfield  {title} {\bibinfo
  {title} {How water droplets evaporate on a superhydrophobic substrate},\
  }\href@noop {} {\bibfield  {journal} {\bibinfo  {journal} {Physical Review
  E}\ }\textbf {\bibinfo {volume} {83}},\ \bibinfo {pages} {026306} (\bibinfo
  {year} {2011})}\BibitemShut {NoStop}%
\bibitem [{\citenamefont {Wilson}\ and\ \citenamefont
  {D'Ambrosio}(2023)}]{wilson2023evaporation}%
  \BibitemOpen
  \bibfield  {author} {\bibinfo {author} {\bibfnamefont {S.~K.}\ \bibnamefont
  {Wilson}}\ and\ \bibinfo {author} {\bibfnamefont {H.-M.}\ \bibnamefont
  {D'Ambrosio}},\ }\bibfield  {title} {\bibinfo {title} {Evaporation of sessile
  droplets},\ }\href@noop {} {\bibfield  {journal} {\bibinfo  {journal} {Annual
  Review of Fluid Mechanics}\ }\textbf {\bibinfo {volume} {55}},\ \bibinfo
  {pages} {481} (\bibinfo {year} {2023})}\BibitemShut {NoStop}%
\bibitem [{\citenamefont {Irving}\ and\ \citenamefont
  {Kirkwood}(1950)}]{irving1950statistical}%
  \BibitemOpen
  \bibfield  {author} {\bibinfo {author} {\bibfnamefont {J.}~\bibnamefont
  {Irving}}\ and\ \bibinfo {author} {\bibfnamefont {J.~G.}\ \bibnamefont
  {Kirkwood}},\ }\bibfield  {title} {\bibinfo {title} {The statistical
  mechanical theory of transport processes. iv. the equations of
  hydrodynamics},\ }\href@noop {} {\bibfield  {journal} {\bibinfo  {journal}
  {The Journal of chemical physics}\ }\textbf {\bibinfo {volume} {18}},\
  \bibinfo {pages} {817} (\bibinfo {year} {1950})}\BibitemShut {NoStop}%
\bibitem [{\citenamefont {Mulero}\ \emph {et~al.}(2009)\citenamefont {Mulero},
  \citenamefont {Cachadina},\ and\ \citenamefont
  {Solana}}]{mulero2009equation}%
  \BibitemOpen
  \bibfield  {author} {\bibinfo {author} {\bibfnamefont {A.}~\bibnamefont
  {Mulero}}, \bibinfo {author} {\bibfnamefont {I.}~\bibnamefont {Cachadina}},\
  and\ \bibinfo {author} {\bibfnamefont {J.}~\bibnamefont {Solana}},\
  }\bibfield  {title} {\bibinfo {title} {The equation of state of the hard-disc
  fluid revisited},\ }\href@noop {} {\bibfield  {journal} {\bibinfo  {journal}
  {Molecular Physics}\ }\textbf {\bibinfo {volume} {107}},\ \bibinfo {pages}
  {1457} (\bibinfo {year} {2009})}\BibitemShut {NoStop}%
\end{thebibliography}

%

\clearpage
\newpage

\appendix

\section{General solution for rational equation of state}\label{Appendix:generalsolution}
Starting from the general constitutive equation \eqref{Eq:const_eq} and a general compressibility term $Z(\psi) = 1/\left( 1- \psi \right)^{\alpha}$, equation \eqref{Eq:const_eq} can be rewritten:
\begin{multline}
    \mathrm{Pe}_{\alpha} \frac{\phi(R)}{\phi(R)-\phi_0}  \tilde{u} =
    2^{\alpha} \left.\frac{ Z}{1-\psi^2} \right[ \left(\alpha -1 \right)\left(1 + \psi \right) +  \\ \left. \frac{2 \left(\phi_m+\left(\alpha -1 \right)\phi_0 \right)}{\phi_m - \phi_0}\right]\nabla' \psi
    \label{Eq:const_equ_gen}
\end{multline}
where $\mathrm{Pe}_{\alpha} = R/\xi_{\alpha,n}$ with $\xi_{\alpha} = \left( \frac{\phi_m}{\phi_m-\phi_0} \right)^{\alpha} D_0/\dot{R}$.
$\tilde{u} = \left( R /r \right)^{n-1}$, with $n$ being the number of system dimensions. \newline
Using simple elements decomposition, it can be shown for any $\alpha$ that:
\begin{equation}
    \frac{Z}{1-\psi^2} = \frac{1}{2^{\alpha}} \left[\frac{1}{1-\psi^2}  + \sum^{\alpha - 1}_{k=0} \frac{2^{k}}{\left( 1-\psi \right)^{k+2}}\right]
\end{equation}
Supposing $\phi(R) \approx \phi_m$,this gives for any $\alpha$ a general solution for the constitutive equation:
\begin{multline}
    \frac{\mathrm{Pe}_{\alpha}}{2} \tilde{u} \mathrm{d}(r') =  2^{\alpha-1} \mathrm{d}\left( \frac{1}{1-\psi}\right)^{\alpha} + \left[ 1+ \left(\alpha-1 \right)\frac{\phi_0}{\phi_m}\right] \\ \times\left\{ \mathrm{d}\left( \arctanh(\psi) \right) + \sum^{\alpha - 1}_{k=1} \frac{2^{k-1}}{k} \mathrm{d}\left(\frac{1}{ 1-\psi }\right)^{k} \right\} 
\end{multline}

\begin{figure}[b]
     \includegraphics[width=.76\columnwidth]{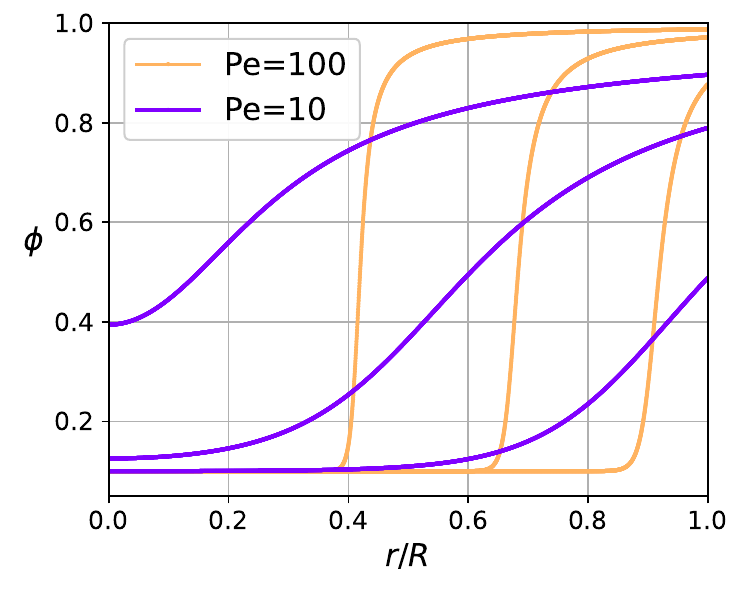}
    \caption{Numerical concentration profiles for $\alpha = 1$ and circular geometry with $\phi_m = 1.0$ and $\phi_0 = 0.1$ obtained by solving equation \eqref{Eq:Gen_eq} using the finite element method described here \cite{diddens2017detailed}.}
\label{fig:Pe_simu}
\end{figure}

\begin{figure*}
    \includegraphics[width=1.6\columnwidth]{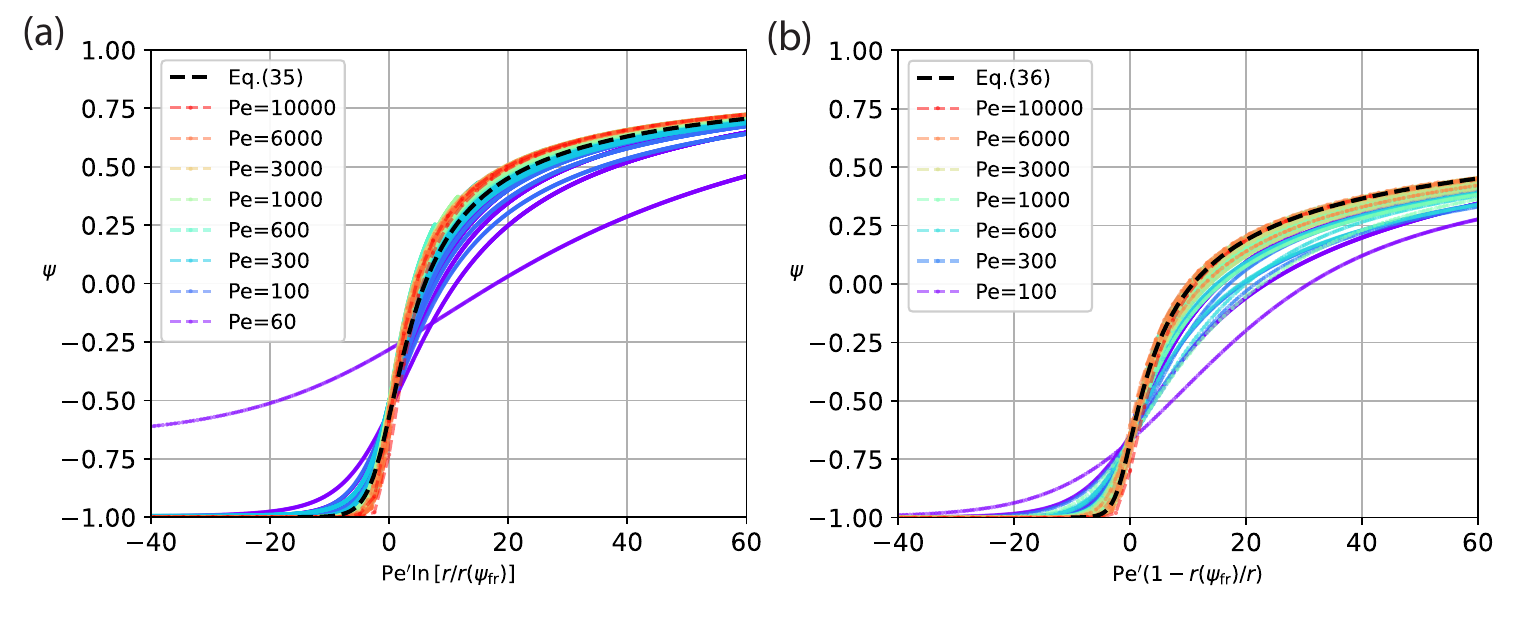}            
    \caption{Numerical concentration profiles for (a) $\alpha = 2$ in circular geometry and (b) and $\alpha = 3$ in spherical geometry in the late regime $h>2R/\mathrm{Pe'}$ and $t<0.95\tau_D$ compared respectively to the quasi-static analytical expressions Eq.\eqref{Eq:An_sol_2D} with $\psi_{\mathrm{fr}} = -0.57$ and Eq.\eqref{Eq:An_sol_3D} with $\psi_{\mathrm{fr}} = -0.68$. $\phi_0$ varies in the range $[0.01,0.8]$ and $\phi_m$ in $[0.4,1.0]$. }
\label{fig:Peresc_simu}
\end{figure*}

Figure \ref{fig:Peresc_simu} shows the numerical solutions for the cases $\alpha = 2$ and $3$ and respectively for circular and spherical geometries in the shell-growing regime ($h>2R/\mathrm{Pe'}$) compared to the asymptotic regime of small $\phi_0 / \phi_m$ of their respective analytical solutions:  
We show here for circular and spherical geometry, the cases $\alpha = 2, 3$, corresponding to the diverging exponent of the Solana equation for hard disks and Carnahan-Starling for spheres:
\begin{multline}
    \alpha = 2: \\ \frac{\mathrm{Pe}_{2}}{2} \ln{\left(\frac{r}{r(\psi_{\mathrm{fr}})}\right)} =  \left[  2\left( \frac{1}{1-\psi}\right)^{2} + \left( 1+\frac{\phi_0}{\phi_m} \right)\times \right. \\  \left. \left\lbrace \arctanh(\psi)+   \frac{1}{1-\psi} \right\rbrace \right]^{\psi}_{\psi_{\mathrm{fr}}}
    \label{Eq:An_sol_2D}
\end{multline}
\begin{multline}
    \alpha = 3: \\ \frac{\mathrm{Pe}_{3}}{2} \frac{R}{r(\psi_{\mathrm{fr}})}\left( 1- \frac{r(\psi_{\mathrm{fr}})}{r}\right) =  \left[ 4\left( \frac{1}{1-\psi}\right)^{3} +  \left( 1+2\frac{\phi_0}{\phi_m} \right)\times \right. \\ \left.\left\lbrace  \arctanh(\psi)+   \frac{1}{1-\psi}  +  \left( \frac{1}{1-\psi}\right)^{2} \right\rbrace \right]^{\psi}_{\psi_{\mathrm{fr}}}
    \label{Eq:An_sol_3D}
\end{multline}
Where $r(\psi_{\mathrm{fr}})$ is the maximum gradient position of the concentration profile given by $\Delta \psi (\psi_{\mathrm{fr}}) = 0$. It gives respectively: 
\begin{align*}
&\text{ for } \alpha = 2 \\
&    \psi_{\mathrm{fr}} = -1 - \frac{4}{3} \frac{\phi_m + \phi_0}{\phi_m - \phi_0}\left( 1- \sqrt{1+ \frac{3}{4} \frac{\phi_m - \phi_0}{\phi_m + \phi_0} } \right)\\
 &\text{ for } \alpha = 3 \\
&    \psi_{\mathrm{fr}} =  -1-\frac{5}{8} \frac{\phi_m + 2\phi_0}{\phi_m - \phi_0}\left( 1- \sqrt{1+ \frac{1}{2}\frac{8^2}{5^2} \frac{\phi_m - \phi_0}{\phi_m + 2\phi_0} } \right) 
\end{align*}

For the spherical case, $R/r(\psi_{\mathrm{fr}})$ is obtained by considering $r(\psi_{\mathrm{fr}}) = R- h$ and $h$ is obtained from \eqref{Eq:h}.
The $\alpha = 2$ solution is used to compare with the Brownian dynamics simulation.

\begin{figure}[h!]
    \includegraphics[width=.9\columnwidth]{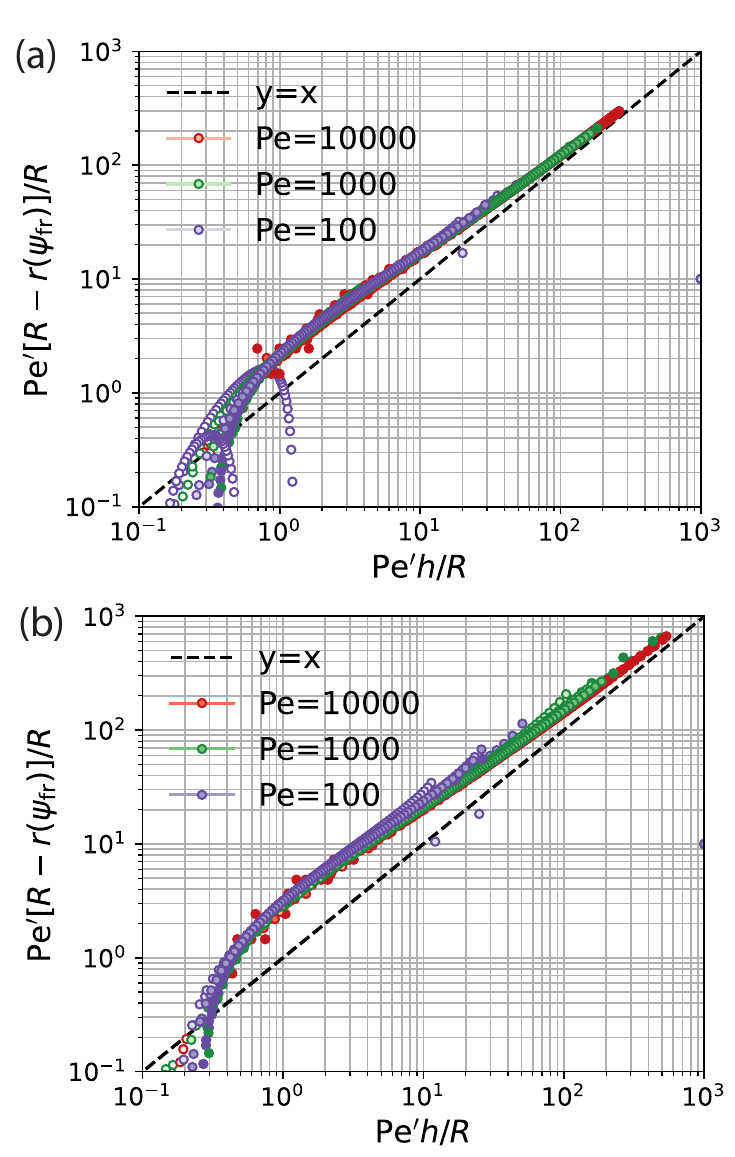}
    \caption{ Maximum gradient position for simulated profiles with (a) $\alpha = 2$ in circular geometry and (b) $\alpha = 3$ in spherical geometry, compared to the shell thickness $h$ derived from expression \eqref{Eq:h}. $\phi_0$ varies in the range $[0.01,0.8]$ and $\phi_m$ in $[0.4,1.0]$. Darker colors correspond to increasing values of $(\phi_m-\phi_0)/\phi_m$.}
\label{fig:h_simu}
\end{figure}

\section{Comparison of the analytical solution with a general numerical solution}
\label{Appendix:comparison}
We compare the previous quasi-static analytical expressions to the finite-element numerical solution of the diffusion equation \eqref{Eq:Gen_eq} in a shrinking domain. The calculation is performed with an in-house finite element framework which is developed in detail on the following references \cite{diddens2017detailed,diddens2017evaporating}. The boundary conditions are a Neumann condition at the domain center and a volume conserving fitted flux at the domain boundary \eqref{Eq:kin_BC}.

After confirming that the fitted flux agrees with the expected boundary condition \eqref{Eq:kin_BC}, a first look at the derived solution Figure \ref{fig:Pe_simu} shows that a quasi-stationary regime is hard to observe for $\mathrm{Pe}<30$. This strong constraint is surprising as the necessary condition to observe a quasi-static profile is only that $\mathrm{Pe} \ll 1$.

We also compare Figure \ref{fig:h_simu} the obtained shell thickness defined here at the position at which $\psi = \psi_{\mathrm{fr}}$ with expression \eqref{Eq:h}. We see that the evolution front position obtained by the numerical calculation shows two regimes. The first one, where the simulated front position does not follow expression \eqref{Eq:h} correspond to the building of the condensed region and typically stops at $ h/R > 1/\mathrm{Pe}_{\alpha}$. The second regime corresponding to the growing of the shell reaches asymptotically expression \eqref{Eq:h}. Typically this point is reached typically for $ h/R > 2^{\alpha} 10/\mathrm{Pe}_{\alpha}$. This also means that for most of the growing shell regime, expression \eqref{Eq:h} tends to underestimate down to a factor 2 to 4 the simulated shell thickness.

For $h>2R/\mathrm{Pe}_{\alpha}$, we show Figure \ref{fig:Peresc_simu} the numerical solution in the shell growing regime for the $\alpha = 2$ and $3$ respectively in the circular and spherical geometry, compared to the analytical solution Eq.\eqref{Eq:An_sol_2D} and Eq.\eqref{Eq:An_sol_3D} in the asymptotic regime of small $\phi_0 / \phi_m$. We observe an excellent agreement for $\mathrm{Pe}>2^{\alpha} \times 10 $, showing that despite the quasi-static approximation is correct on the reference frame of the moving front (given here by $r(\psi_{\mathrm{fr}})$). It also shows that the shell front lengthscale is correctly given by $\xi_{\alpha, n}$. We observe that in the parameter range shown, the value $\psi =+1$ is not approached showing the effect of the $\alpha$ parameter on the compressibility property of the system. This explains the underestimation of equation \eqref{Eq:h} observed for the shell thickness value.

\section{Brownian particle simulation equation of state}
\label{Appendix:Brownian particle sims eq of state?}
We extract the particles pressure of the Brownian particles system by running multiple simulations of $N = 5000$ particles diffusing on fixed domain of area $A$. The total pressure of the system $P$ is calculated from the summation of kinetic energy and the work done by all particles after the particles has reached a thermodynamic equilibrium. For an isotropic and homogeneous hard disks system, according to the virial equation \cite{irving1950statistical}: 
\begin{equation}
    P = \frac{1}{A}\left( N k_B T + \Sigma_{i} \Sigma_{j>i} F_{ji} r_{ji} \right)
    \label{Eq:P}
\end{equation}
where $F_{ji}$ is the pair particles repulsive WCA forces and $r_{ji}$ the inter-particles distance.
Here the particles having no inertia, the thermal energy is converted into viscous work giving $k_B T = \xi_f v^2 \Delta t / 2 $ with $\Delta t = \sigma^2_d / 2 D_0$ the correlation time of the particle displacement and $\sigma_d = v \Delta t$ its corresponding correlation distance. Here $\Delta t$ corresponds to the simulation time step.

\begin{figure}[b]
     \includegraphics[width=\columnwidth]{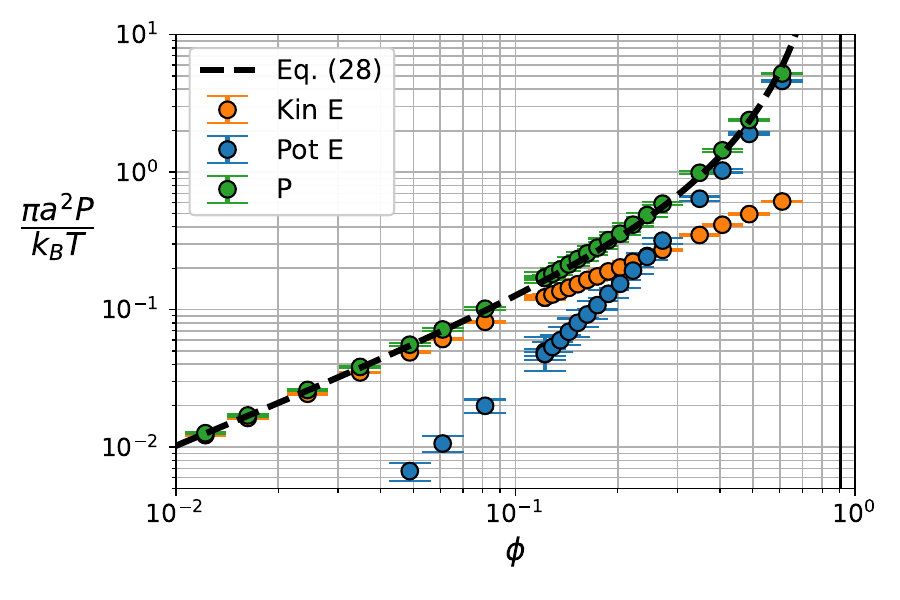}
    \caption{Normalised particles pressure obtained from \eqref{Eq:P} in function of the volume fraction. The black solid line shows the maximum packing fraction $\phi_m =0.9069$.}
\label{fig:Osm_press}
\end{figure}

Using equation \eqref{Eq:P}, we show Figure \ref{fig:Osm_press} that the normalised pressure is given by Solana hard-disk equation of state \cite{mulero2009equation}:
\begin{equation}
    \phi Z(\phi) = \phi \frac{1+\frac{1}{8}\eta^2 - \frac{1}{10}\eta^4}{\left(1- \eta \right)^2}
    \label{Eq:EqS_Solana}
\end{equation}
with $\eta = \phi / \phi_m$ with $\phi_m =0.9069$ the maximum packing fraction for hard disk system. This equation of state only differs by a maximum of $2.5\%$ from using a simple compressibility term \eqref{Eq:Comp_term} with $\alpha = 2$. 

\end{document}